\documentclass[twocolumn,prb,showpacs,preprintnumbers,superscriptaddress,amsmath,amssymb,citeautoscript]{revtex4-1}
\usepackage{graphicx}
\usepackage{epstopdf}
\usepackage{dcolumn}
\usepackage{color}
\usepackage{amsmath,amssymb}
\usepackage[english]{babel}
\usepackage{bm}
\usepackage[hyperindex]{hyperref}
\hypersetup{colorlinks,citecolor=blue, filecolor=blue,linkcolor=blue,urlcolor=blue}

\graphicspath{{./figures/}}
\usepackage{multirow}
\usepackage{rotating}
\usepackage{array}

\newcommand{\bpm}{\begin{pmatrix}}
\newcommand{\epm}{\end{pmatrix}}
\newcommand{\bs}{\boldsymbol}
\newcommand{\be}{\begin{equation}}
\newcommand{\ee}{\end{equation}}
\newcommand{\beq}{\begin{eqnarray}}
\newcommand{\eeq}{\end{eqnarray}}

\DeclareMathOperator{\im}{Im}

\DeclareMathOperator{\tr}{tr}

\begin{document}

\title{Surface Green's functions and quasiparticle interference in Weyl semimetals}

\author{Sarah Pinon}
\affiliation{Institut de Physique Th\'eorique, Universit\'e Paris Saclay, CEA
CNRS, Orme des Merisiers, 91190 Gif-sur-Yvette Cedex, France}
\author{Vardan Kaladzhyan}
\email{vardan.kaladzhyan@phystech.edu}
\affiliation{Department of Physics, KTH Royal Institute of Technology, Stockholm, SE-106 91 Sweden}
\author{Cristina Bena}
\affiliation{Institut de Physique Th\'eorique, Universit\'e Paris Saclay, CEA
CNRS, Orme des Merisiers, 91190 Gif-sur-Yvette Cedex, France}

\date{\today}

\begin{abstract}
We use the exact analytical technique introduced in Phys. Rev. B \textbf{101}, 115405 to recover the surface Green's functions and the corresponding Fermi-arc surface states for various lattice models of Weyl semimetals. For these models we use the T-matrix formalism to calculate the quasiparticle interference patterns generated in the presence of impurity scattering. In particular, we consider the models introduced in Phys. Rev. B \textbf{93}, 041109(R) (A) and Phys. Rev. Lett. \textbf{119}, 076801 (B), and we find that, as opposed to observations previously obtained via joint density of states and spin-dependent scattering probability, the inter-arc scattering in the quasiparticle interference features is fully suppressed in model A, and very small in model B. Our findings indicate that these models may not correctly describe materials such as  MoTe$_2$, since for such materials inter-arc scattering is clearly visible experimentally, e.g., in Nature Phys. \textbf{12}, 1105–1110 (2016). We also focus on the minimal models proposed by McCormick et al. in Phys. Rev. B \textbf{95}, 075133, which indeed recover significant inter-arc scattering features. 

\end{abstract}

\maketitle

\section{Introduction}

Weyl semimetals\cite{Turner2013} have come recently into focus partly due to their exotic properties of exhibiting Weyl nodes, i.e., points at which the energy dispersion goes to zero and quasiparticles have a linear dispersion, as well as peculiar topological surface states, deemed Fermi-arc states. Typically these states appear at values of in-plane momentum lying on a line connecting the Weyl nodes, but for some of them the picture is modified by the appearance of surface electron/hole pockets; for these type of systems other non-topological surface states also arise. Such states, denoted track states, appear in momentum space as closed contours (as opposed to the Fermi-arc states). 

A large amount of work, both theoretical\cite{Mitchell2016,Gyenis2016,Lambert2016,Chang2016,Kourtis2016,McCormick2017,Lau2017,Xu2018,Zheng2018} and experimental\cite{Jeon2014,Deng2016,Zheng2016a,Zheng2016b,Inoue2016,Batabyal2016,Zhang2017,Deng2017,Lin2017,Yuan2018,Lin2018,Morali2019,Sessi2020}, has been devoted to the impurity scattering and the corresponding quasiparticle interference (QPI) features, and the possibility to extract information about the Weyl nodes, and hence Fermi arcs, from such features. However, the analysis of the QPI features is very complex, especially in the presence of both topological and trivial states simultaneously. The traditional methods to obtain the QPI features stem in general from first principle calculations yielding information about the position, shape and extension of Fermi arcs and track states in momentum space. Often a simple intuitive picture is constructed which assumes that the QPI features stem directly from a joint density of states (JDOS) of the underlying Fermi-arc states: in this picture each surface state is assumed to scatter by all the other existing surface states, and the resulting scattering momenta are counted to yield the QPI feature. The first correction to this picture takes into account the spin of each state and introduces a weight to the scattering processes based on the spins of each scattering pair of states. However, these techniques are very crude and often fail to capture correctly the scattering processes and to predict the QPI features accurately. Their advantage, nevertheless, is that only basic information about the Fermi-arc states is required.

In order to calculate accurately the QPI features a more complex technique---the T-matrix formalism\cite{Ziegler1996,Salkola1996,Mahan2000,Bena2016}---should be employed. The latter allows to calculate exactly the Green's function of the system in the presence of a single impurity, and thus enables to obtain the local density of states, as well as its Fourier transform, in the presence of impurity scattering. However, in order to apply the T-matrix formalism, information about the underlying surface Green's function of the system is required. The surface Green's functions of a given system cannot be computed very easily, in general it requires  iterative numerical techniques, hard to implement in a non-numerical community. 

In a recent work\cite{Pinon2020} we proposed a new simple and exact analytical technique that allows to obtain the surface Green's functions of a three-dimensional system, with the only necessary ingredient being the bulk tight-Hamiltonian of the system. To summarize, the idea is to start with an infinite homogeneous three-dimensional system and then introduce the boundary as a two-dimensional planar scalar impurity with an impurity potential much larger than all the other energy scales of the problem \footnote{Note that the impurity needs to contain a number of planes  equal to the unit cell number of planes in the model, or more in case of higher-order hopping between planes}. This impurity effectively cuts the system in half and generates two two-dimensional infinite surfaces on the planes just below and above the impurity plane. The surface Green's functions for these surfaces are obtained using the T-matrix formalism, which for such a configuration provides an exact analytical solution for the full Green's functions of the infinite perturbed system, and thus also for the Green's function evaluated at positions adjacent to the impurity plane.

In this work we use the technique from Ref.~[\onlinecite{Pinon2020}] to calculate the surface Green's functions for a number of models proposed to describe Weyl semimetals: 1) the Kourtis\cite{Kourtis2016} and Lau\cite{Lau2017} models, minimal models that describe mostly type I Weyl semimetals with four Weyl points and two Fermi arcs, and 2) the McCormick model\cite{McCormick2017} based on a minimal Helium model to describe both type I and type II four-Weyl-points systems with both Fermi arcs and topologically trivial track states. We subsequently use the resulting surface Green's functions and the T-matrix formalism to calculate the QPI features for these systems in the presence of a localized impurity at the surface. We find that in the Kourtis and Lau models the inter-arc scattering is almost entirely absent as a result of the spin structure of the Fermi-arc states. This is different from the results based on spin-dependent scattering probability (SSP) calculations in which the spin structure of the Fermi arcs reduces, but does not completely remove, the inter-arc features. As for the McCormick model, we identify features resulting from intra- and inter-Fermi-arc scattering,  intra- and inter-track-states scattering as well as scattering between the Fermi arc states and the track states. 

The paper is organized as follows: in Sec.~II we present the Kourtis and Lau, models, their surface GF and the corresponding QPI patterns. In Sec.~III we perform similar calculations for a model McCormick model, and we conclude in Sec.~IV.

\section{Surface Green's functions and QPI patterns in the Kourtis and Lau minimal models for MoTe$_2$}

We first consider the minimal models introduced in Refs.~[\onlinecite{Kourtis2016}] and [\onlinecite{Lau2017}] to describe systems in the family of MoTe$_2$, i.e., Weyl semimetals with four Weyl points and two Fermi arcs per surface. The surface Green's functions and Fermi arcs, as well as their spin polarization have been obtained by the analytical technique presented in Ref.~[\onlinecite{Pinon2020}], thus below we summarize these findings, and proceed directly to computing the corresponding QPI features.

The tight-binding models described in Refs.~[\onlinecite{Kourtis2016}] and [\onlinecite{Lau2017}], which we denote by $\mathcal{H}_1$ and $\mathcal{H}_2$, respectively, are described by the Bloch Hamiltonians given by
\begin{eqnarray}
H_{1,2}=\sum_{\bs{k}} \psi^\dagger(\bs{k}) \mathcal{H}_{1,2}(\bs{k})\psi(\bs{k}),
\end{eqnarray}
where $\psi(\bs{k})=(c_{\bs{k} A \uparrow},c_{\bs{k} A \downarrow}, c_{\bs{k} B \uparrow}, c_{\bs{k} B \downarrow})$ is a spinor with the index $A/B$ denoting a generic unspecified orbital component, and $ \uparrow/ \downarrow$ the physical spin. 

For the model in Ref.~[\onlinecite{Kourtis2016}] written in the aforementioned basis we have
\begin{eqnarray}
\mathcal{H}_{1}(\bs{k})&=&g_1(\bs{k}) \tau_1\sigma_3+g_2(\bs{k}) \tau_2 \sigma_0+g_3(\bs{k})\tau_3\sigma_0
\nonumber\\&&+g_0(\bs{k})\tau_0\sigma_0+\beta \tau_2\sigma_2+\alpha \sin k_y \tau_1\sigma_2,
\end{eqnarray}
where 
\begin{eqnarray}
g_0(\bs{k})&=&2d(2-\cos k_x-\cos k_y)\nonumber \\
g_1(\bs{k})&=&a \sin k_x\nonumber \\
g_2(\bs{k})&=&a \sin k_y\nonumber \\
g_3(\bs{k})&=&m + t \cos k_z+ 2b(2-\cos k_x-\cos k_y),
\end{eqnarray}
and $\alpha$, $\beta$ are real parameters. The $2\times 2$ identity matrices $\sigma_0/\tau_0$ and the Pauli matrices $\sigma_i/\tau_i$, $i=1,2,3$, act in the spin and the orbital spaces respectively, and side-by-side appearances of $\sigma$ and $\tau$ matrices indicate tensor products.

\begin{figure}
\centering
\hspace*{-0.5cm}
\includegraphics[width=9.6cm]{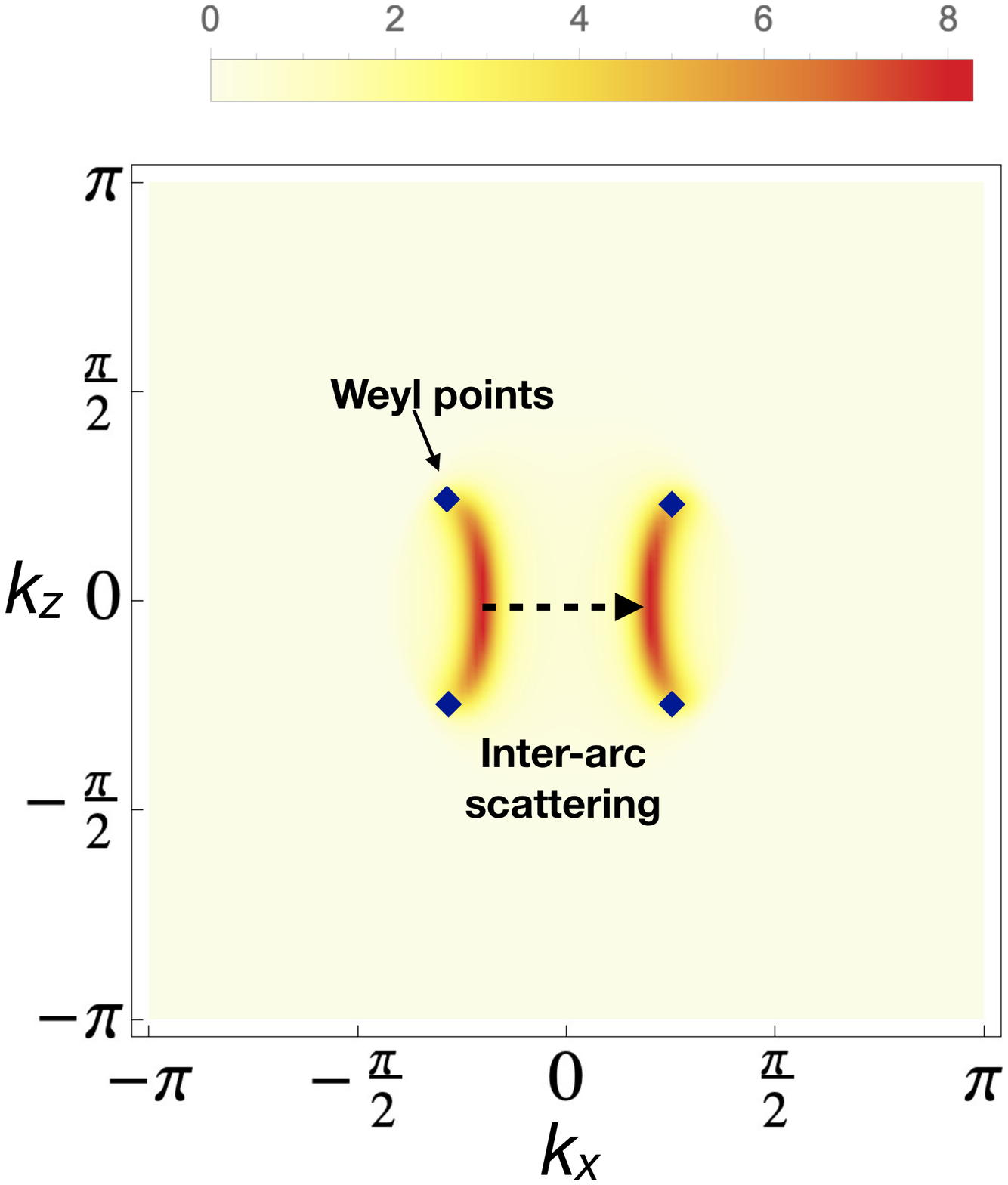}
\caption{The surface spectral function at $E=0$ for the $\mathcal{H}_1$ model with parameters  $a = b = 1$, $t = -1.5$, $d = m = 0$, $\beta = 0.9$. We clearly see the formation of two Fermi arcs. }
\label{fig:spectral_function_H1_Kourtis}
\end{figure}

We choose the following values of the parameters $a = b = 1$, $t = -1.5$, $d = m = 0$, $\beta = 0.9$, and $\alpha = 0.3$, corresponding to four Weyl points, and two Fermi arcs. The spectral function for the surface states $A(E, k_x, k_z) = -\frac{1}{\pi} \im\{\tr[G_s\left(E, k_x, k_z\right) ]\}$, as obtained in Ref.~[\onlinecite{Pinon2020}], is shown in Fig.~\ref{fig:spectral_function_H1_Kourtis}.
We should note that Fermi arcs appearing at the same surface have opposite spins, as derived in Ref.~[\onlinecite{Pinon2020}].

In what follows we start with the surface Green's functions for this model and we compute the surface QPI patterns in the presence of a surface impurity. We use the full T-matrix formalism yielding an exact solution to the problem provided the impurity is localized (i.e., delta function potential $V \delta(\bs{r})$, where $\bs{r}$ is in real space).
In this simple case the T-matrix is given by\cite{Ziegler1996,Salkola1996,Mahan2000,Bena2016, Balatsky2006,Bena2008}:
\be
T(E) = \left[\mathbb{I} -V \int \frac{d\bs{k}}{(2\pi)^2} G_s(E,\bs{k})  \right]^{-1} V.
\label{tm0}
\ee
Generally, in most experiments QPI patterns corresponding to the Fourier transform of the local density of states at a given energy are measured, namely,
$\delta\rho(\bs{k},E) = \int d^2 \bs{r} \, \delta\rho(\bs{r},E) e^{-i \bs{k r}} $. In the T-matrix formalism this is given by 
\be
\delta\rho(\bs{k},E)= \frac{i}{2 \pi} \negthickspace\int \negthickspace\frac{d\bs{q}}{(2\pi)^2} \tr\left[g(E,\bs{q},\bs{k})\right]
\label{eq:QPIintegral}
\ee
where $d\bs{q} \equiv dq_x dq_z$ and 
\beq
g(E,\bs{q},\bs{k}) &=& G_s (E,\bs{q}) T(E) G_s(E, \bs{k+q}) \nonumber\\
&-& G^*_s(E, \bs{k+q}) T^*(E) G^*_s (E,\bs{q}). \nonumber
\eeq

\begin{figure}
\centering
\hspace*{-0.5cm}
\includegraphics[width=9.6cm]{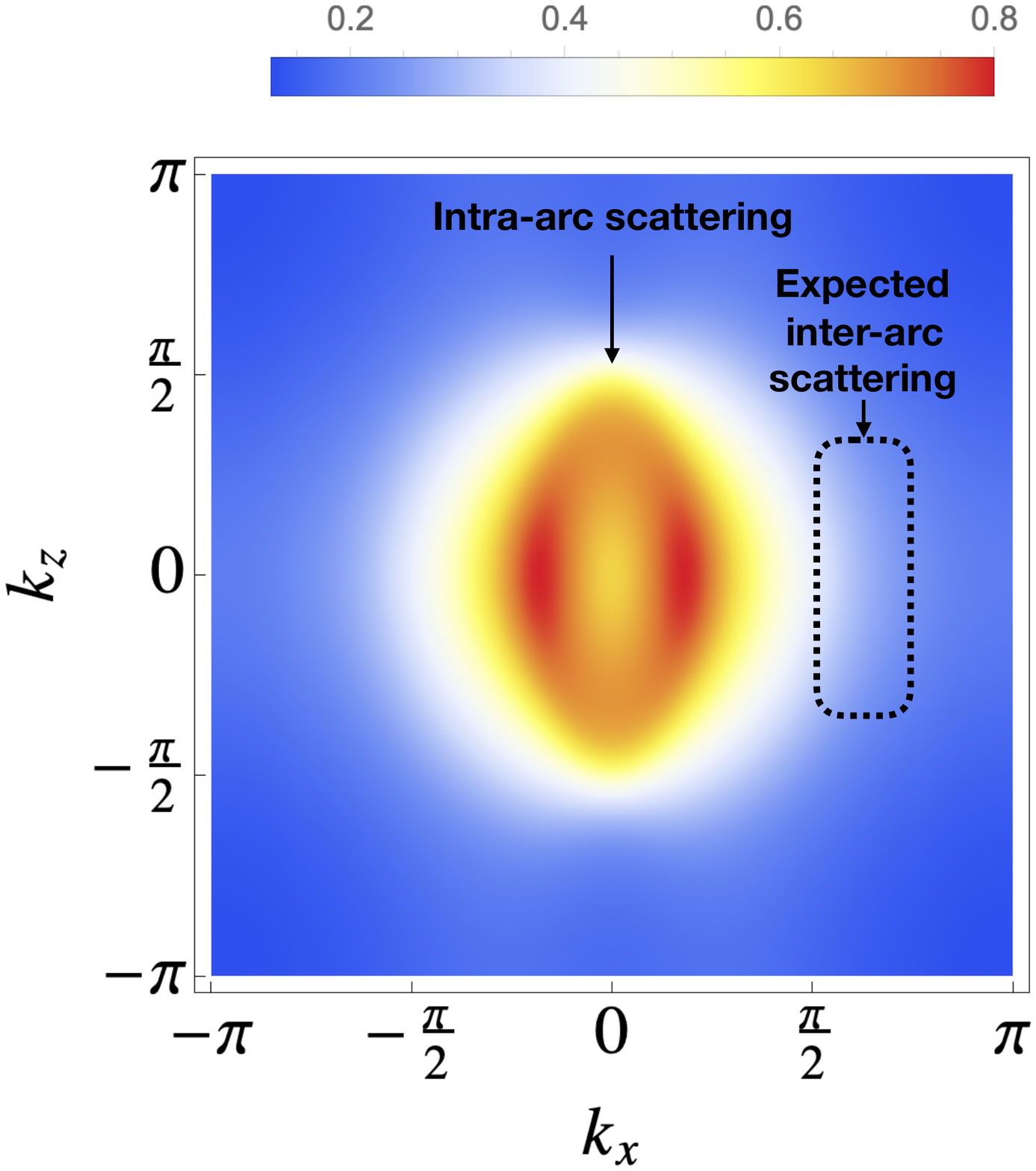}
\caption{QPI patterns for the $\mathcal{H}_1$ model with parameters  $a = b = 1$, $t = -1.5$, $d = m = 0$, $\beta = 0.9$. We take $E=0$.}
\label{fig:QPI_H1_Kourtis}
\end{figure}

In Fig.~\ref{fig:QPI_H1_Kourtis} we plot the resulting $\delta\rho(\bs{k},E)$ as a function of momenta $\bs{k} = (k_x, k_z)$ at $E=0$. First, note the central feature corresponding to intra-arc scattering processes, and second --- the complete absence of any noncentral feature that could potentially arise from the inter-arc scattering processes. Given the spin structure of the two oppositely polarized Fermi arcs, such inter-arc scattering processes should indeed be forbidden, since that would require spin flips which cannot occur in the presence of a scalar scattering potential. For instance, in superconductors, scattering between states with opposite spins does not yield any QPI features\cite{Kaladzhyan2016}. It is surprising that SSP calculations in Ref.~[\onlinecite{Kourtis2016}] indicate, however, a diminished but non-zero intensity in the inter-arc scattering features: as we show here, the full T-matrix approach seems to indicate that the inter-arc scattering is, in fact, zero. This difference probably stems from SSP yielding inaccurate results versus the T-matrix formalism\cite{Derry2015}, since the latter takes into account all-order scattering processes, unlike the former. The only nonzero observable intensity would appear in our approach in the presence of a magnetic impurity in the spin-polarized LDOS, same as in Ref.~[\onlinecite{Kaladzhyan2016}]. However, the present experiments do not measure the spin-polarized LDOS, and they observe nevertheless inter-arc scattering features. This makes us believe that the Hamiltonian for MoTe$_2$ presented in Ref.~[\onlinecite{Kourtis2016}] does not model well the underlying spin structure.
Furthermore, the elliptical shape of the central feature in Fig.~\ref{fig:QPI_H1_Kourtis} is very different from the eight-shaped one derived using SSP and JDOS.

\begin{figure}
\centering
\hspace*{-0.5cm}
\includegraphics[width=9.6cm]{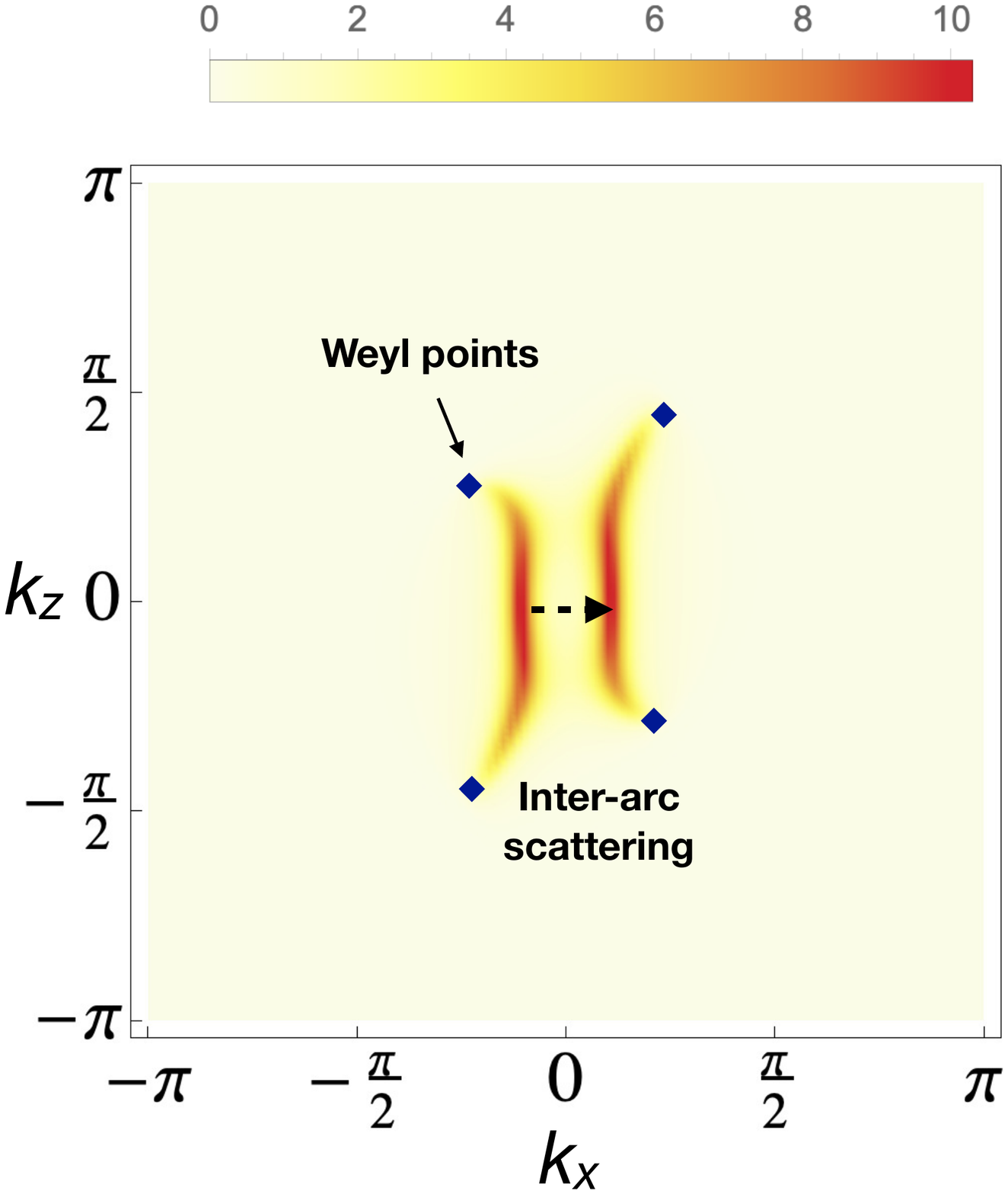}
\caption{The surface spectral function at $E=0$ for the $\mathcal{H}_2$ model with parameters  $a = b = 1$, $t = -1.5$, $\lambda = 0.5$, $d = 0.1$, $\alpha=0.3$ and $\beta=0.7$. We note the emergence of the two Fermi-arcs.  }
\label{fig:spectral_function_H2_Lau}
\end{figure}

We perform a similar analysis on a different Weyl semimetal model, introduced in Ref.~[\onlinecite{Lau2017}]: 
\begin{eqnarray}
\mathcal{H}_{2}(\bs{k})&=&g_1(\bs{k}) \tau_1\sigma_3+g_2(\bs{k}) \tau_2 \sigma_0+g_3(\bs{k})\tau_3\sigma_0+d \tau_2\sigma_3
\nonumber\\&&+\beta \tau_2\sigma_2+\alpha \sin k_y \tau_1\sigma_2+\lambda \sin k_z \tau_0 \sigma_1
\end{eqnarray}
We consider the values of the parameters similar to those in Ref.~[\onlinecite{Lau2017}], thus we take $a = b = 1$, $t = -1.5$, $\lambda = 0.5$, $d = 0.1$, $\alpha=0.3$ and $\beta=0.7$. This configuration is also characterized by four Weyl points, and two Fermi arcs per surface. The resulting surface spectral function is depicted in Fig.~\ref{fig:spectral_function_H2_Lau}.

\begin{figure}
\centering
\hspace*{-0.5cm}
\includegraphics[width=9.6cm]{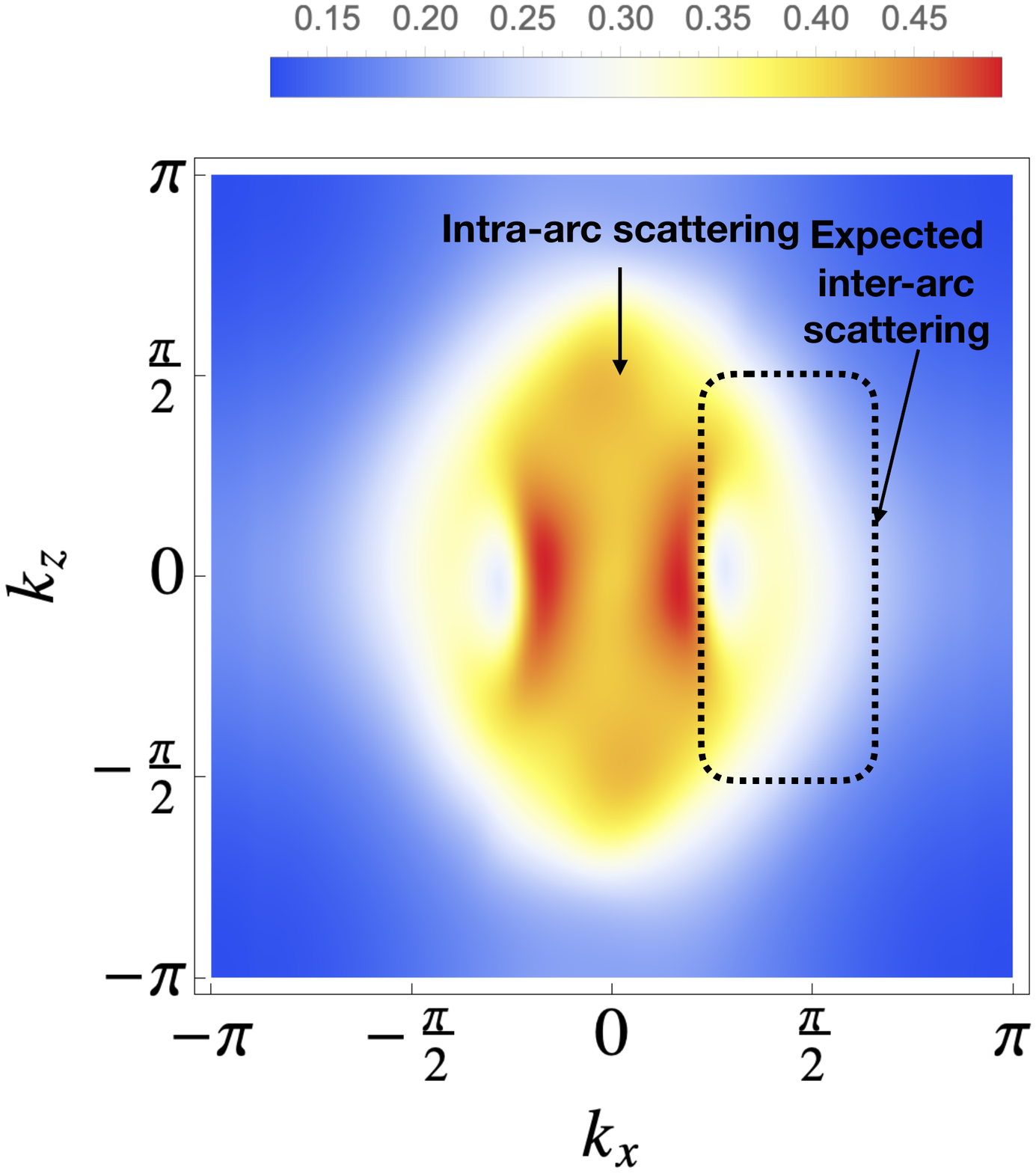}
\caption{QPI patterns for the model $\mathcal{H}_2$ with  $a = b = 1$, $t = -1.5$, $\lambda = 0.5$, $d = 0.1$, $\alpha=0.3$ and $\beta=0.7$. We take $E=0$.}
\label{fig:QPI_H2_Lau}
\end{figure}

For this model we also calculate the QPI patterns, shown in Fig.~\ref{fig:QPI_H2_Lau}. Unlike in the Kourtis model\cite{Kourtis2016}, here the Fermi arcs have \textit{nearly} opposite spins (the two opposing Fermi arcs may exhibit small regions with the same spin as can be seen for example in Fig.~5 of Ref.~[\onlinecite{Pinon2020}]), and therefore, weak inter-arc scattering processes marked by a dotted rectangle are visible in Fig.~\ref{fig:QPI_H2_Lau}. We note, however, that the intra-arc scattering still dominates in the QPI pattern. 

\section{Surface Green's functions and QPI patterns in the model of McCormick et al.}
\begin{figure}
\centering
\includegraphics[width=9cm]{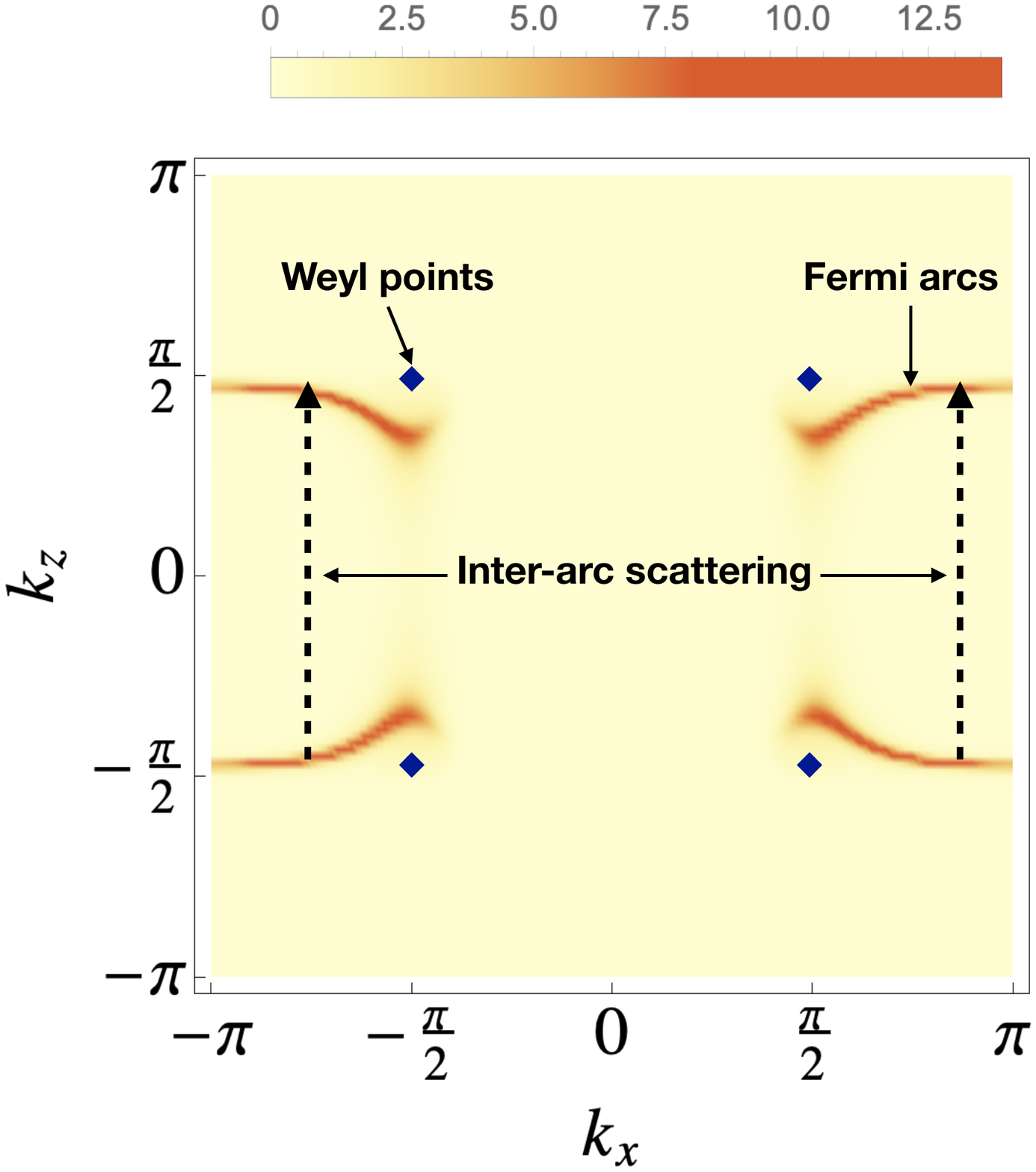}\\
\includegraphics[width=9cm]{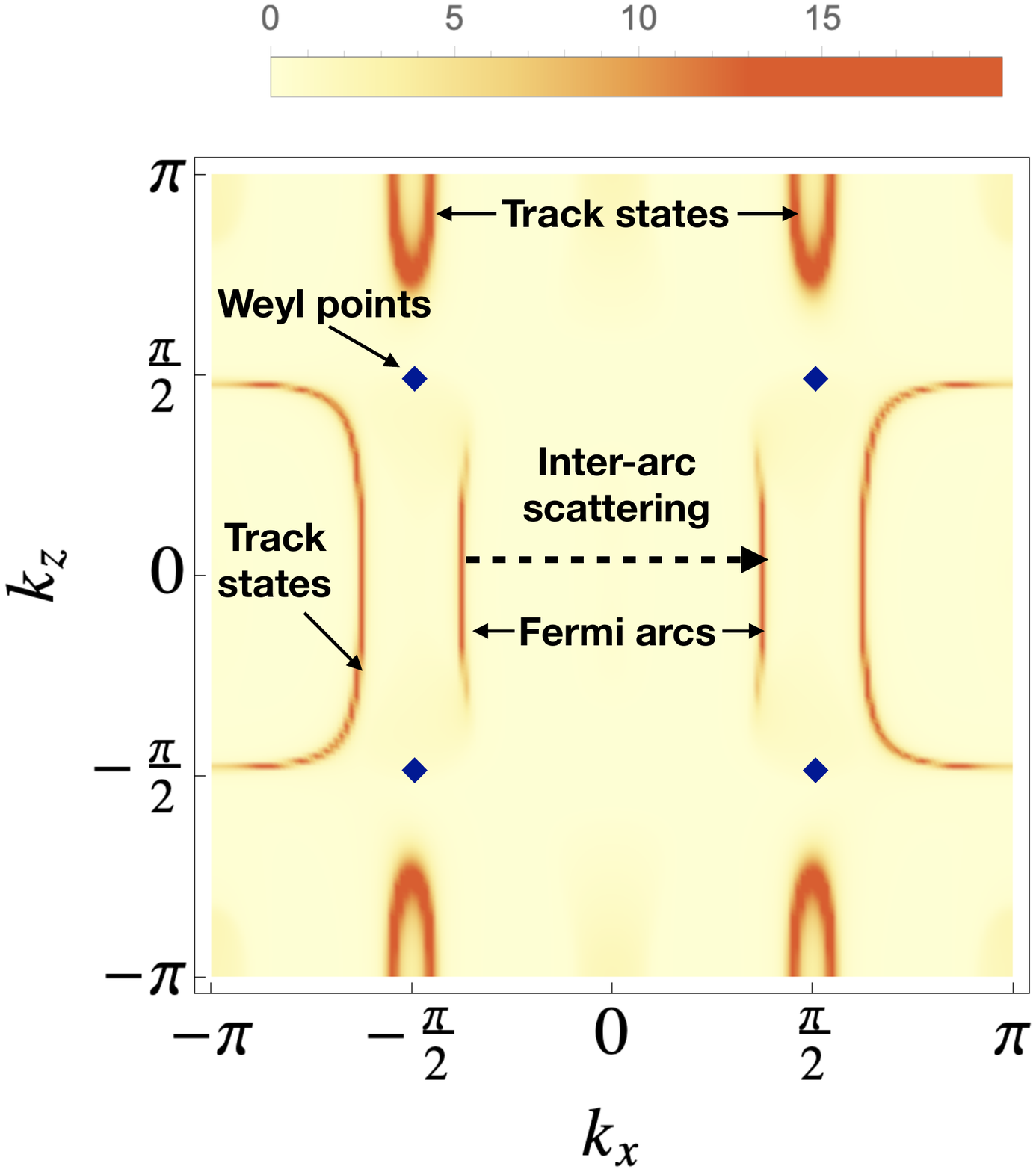}
\caption{The surface spectral function at $E=0.25$ for the $\mathcal{H}_3$ model with parameters  a) $t=1$, $k_0 = \pi/2$, $t_x = 1/2$, $m = 2$, $\gamma=1.4$, and b) $t=1$, $k_0 = \pi/2$, $t_x = 1/2$, $m = 2$, $\gamma=2.4$. }
\label{fig:spectral_function_H3_McCormick}
\end{figure}

In what follows we consider the minimal Helium model proposed in Ref.~[\onlinecite{McCormick2017}] to describe systems with four Weyl nodes, corresponding Fermi-arc surface states, impurity scattering and the resulting QPI patterns. The Hamiltonian for such a model is given by:
\begin{eqnarray}
\mathcal{H}_{3}(\bs{k})&=&\gamma [\cos(2 k_x)-\cos (k_0)][\cos(k_z)-\cos(k_0)]\sigma_0
\nonumber\\&&
-m\{[1-\cos^2(k_z)-\cos(k_y)]\nonumber\\&&+2 t_x[\cos(k_x)-\cos(k_0)]\}\sigma_1
\nonumber \\ &&
-2 t \sin(k_y) \sigma_2- 2 t \cos(k_z) \sigma_3.
\end{eqnarray}
As opposed to the previous section, this model is defined in a single-orbital basis, $\psi(\bs{k})=(c_{\bs{k} \uparrow},c_{\bs{k} \downarrow})$. 
At $\gamma = 0$ it yields a Weyl semimetal with four nodes located at  $(\pm k_0,0,\pm
\pi/2)$. A nonzero $\gamma$ produces a tilting of the Weyl cones \cite{}; when $\gamma$ is
sufficiently large (here $\gamma_c=2$), the modification of the Weyl cones is strong enough to generate four electron and four hole pockets that exist at $E = 0$ and meet at the Weyl nodes. There is also a trivial hole pocket centered at $k = (0,0,0)$ and a trivial electron pocket centered
at $k = (\pi,0,0)$. In this type of systems some of the resulting surface states are topological, i.e, Fermi arcs, while the others, denoted ``track states'',  are trivial. The latter form closed loops in momentum space. We consider two sets of parameters, the first a) $t=1$, $k_0 = \pi/2$, $t_x = 1/2$, $m = 2$, $\gamma=1.4$, exhibiting solely Fermi arcs, and the second  b) $t=1$, $k_0 = \pi/2$, $t_x = 1/2$, $m = 2$, $\gamma=2.4$, with both Fermi arcs and track states. For each set we apply the novel technique introduced in Ref.~[\onlinecite{Pinon2020}] to recover the surface states (i.e., Fermi arcs and track states), and subsequently we use the surface GF obtained by this technique to study the QPI patterns. In Fig.~\ref{fig:spectral_function_H3_McCormick} we plot the spectral function for the surface states $A(E, k_x, k_z) = -\frac{1}{\pi} \im\{\tr[G_s\left(E, k_x, k_z\right) ]\}$ for these two sets of parameters.

Focusing on one boundary (corresponding to that denoted in blue in Ref.~[\onlinecite{Pinon2020}]), we note in Fig.~\ref{fig:spectral_function_H3_McCormick}a the formation of two Fermi arcs roughly parallel to $k_x$, and in Fig.~\ref{fig:spectral_function_H3_McCormick}b that of two Fermi arcs in the $k_y$ direction as well as of two sets of track states. Our results are fully consistent with those of Ref.~[\onlinecite{Pinon2020}], showing once more the versatility and simplicity of our method. Note again that our technique is fully analytical, the only ``numerical'' calculation to be performed is an integral of the Green's function over the Brillouin zone.

\begin{figure}
\centering
\hspace*{-0.5cm}
\includegraphics[width=9.6cm]{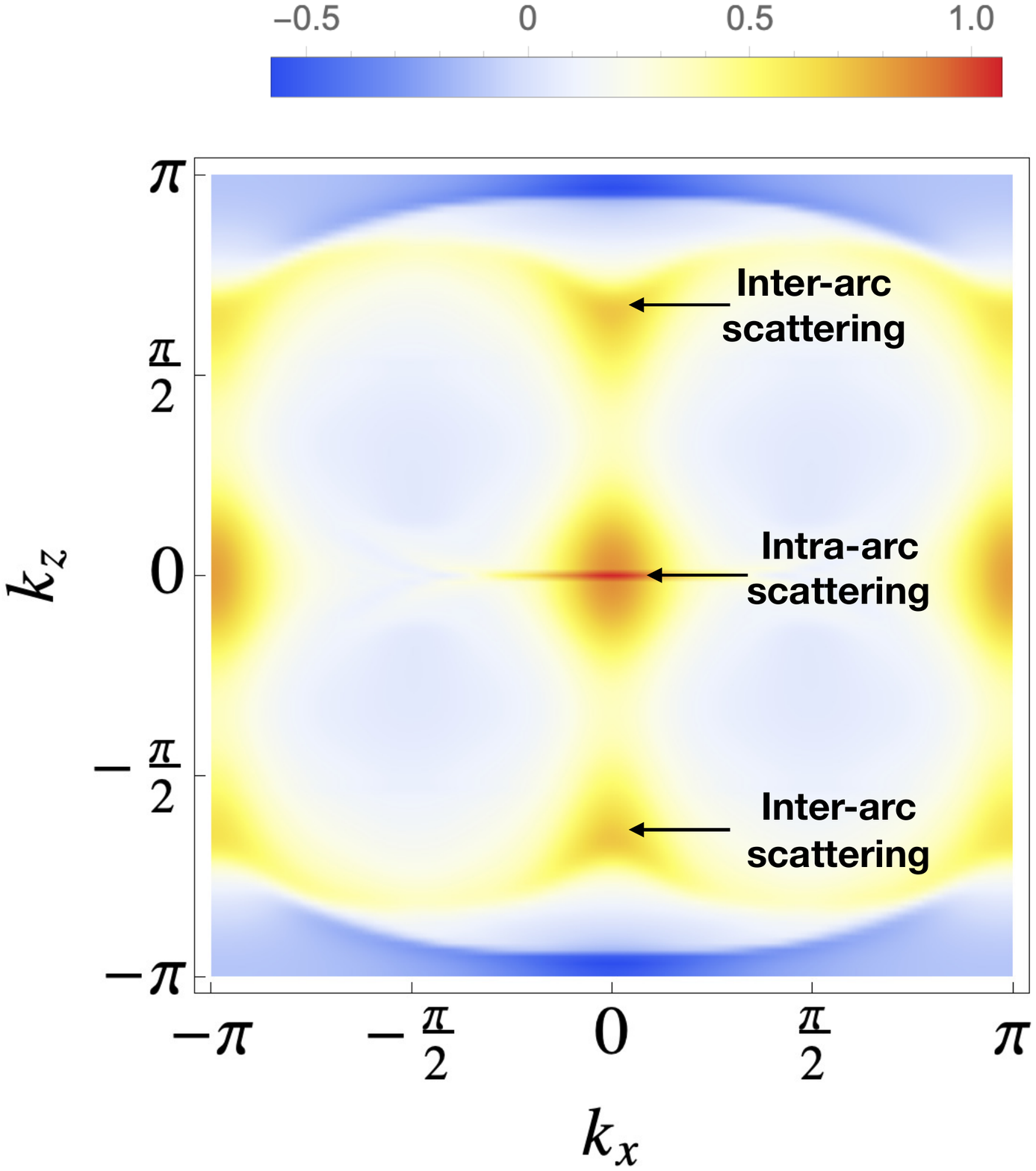}
\caption{QPI patterns for the model $\mathcal{H}_3$ with  a) $t=1$, $k_0 = \pi/2$, $t_x = 1/2$, $m = 2$, $\gamma=1.4$.  We take $E=0.25$.}
\label{fig:QPI_H3_McCormick}
\end{figure}

Once we have recovered the surface Green's functions, we proceed to the calculation of the QPI features in the presence of a single localized impurity. Exactly as in the previous section we use the T-matrix approximation to obtain the Fourier transform of the LDOS, i.e., a quantity that can be compared to what is currently measured experimentally. In Fig.~\ref{fig:QPI_H3_McCormick} we plot the resulting $\delta\rho({\bs k},E)$ as a function of $\bs{k}$ for $E=0.25$ and the set of parameters of Fig.~\ref{fig:spectral_function_H3_McCormick}a. 
We note the presence of a central feature which corresponds to the intra-Fermi-arc scattering, as well as a wing-like feature centered around values of $k_z$ slightly larger in absolute value than $\pi/2$, corresponding to inter-arc scattering.  Note that, as opposed to the previous section, the inter-arc and intra-arc scattering features are of the same order of intensity.

\begin{figure}[b!]
\centering
\vspace*{-0.7cm}
\includegraphics[width=9.6cm]{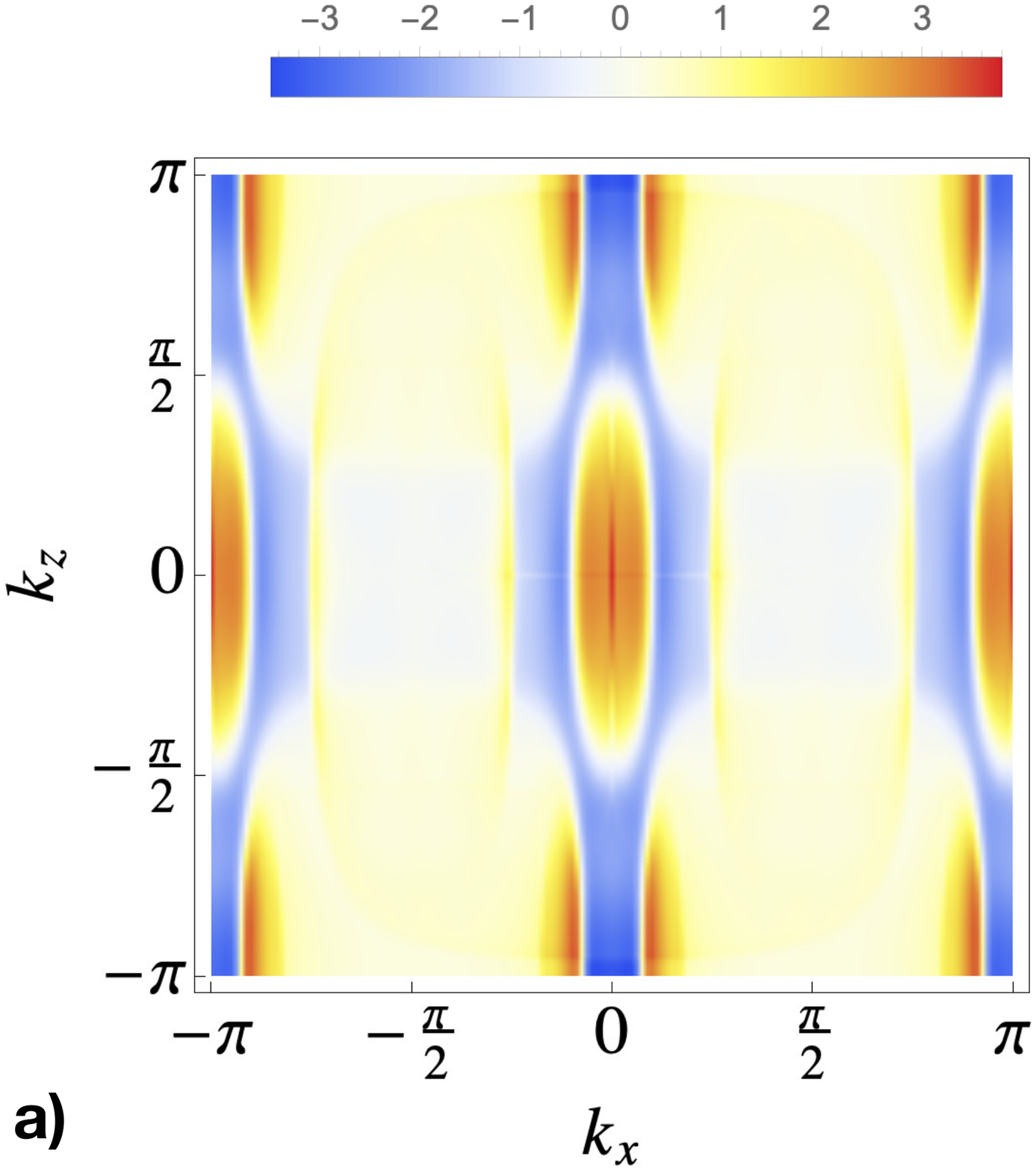}\\
\hspace*{-0.5cm}
\includegraphics[width=9.2cm]{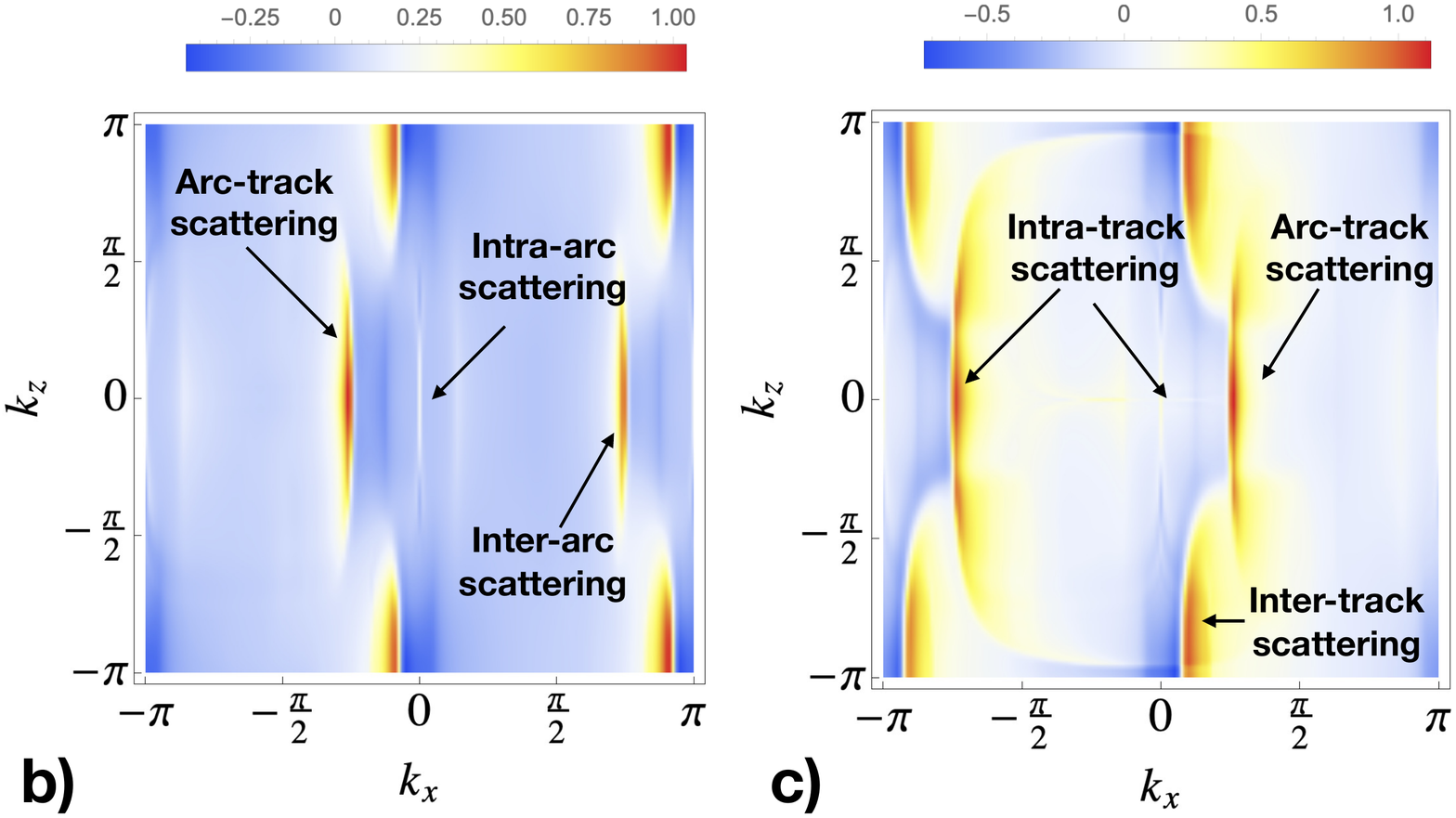}
\vspace*{-1.3cm}
\caption{QPI patterns for the model $\mathcal{H}_3$ with  $t=1$, $k_0 = \pi/2$, $t_x = 1/2$, $m = 2$, $\gamma=2.4$. We take $E=0.25$.}
\label{fig:QPI_track_H3_McCormick}
\end{figure}

In Fig.~\ref{fig:QPI_track_H3_McCormick}a we plot the QPI features observed for the other set of parameters, i.e., $t=1$, $k_0 = \pi/2$, $t_x = 1/2$, $m = 2$, $\gamma=2.4$, corresponding to the formation of both topological Fermi arcs and trivial track states. We note that the resulting QPI features are indeed very complex, since they correspond to both intra- and inter-Fermi-arc scattering, as well scattering inside and between different closed contours representing the track states, and also to the scattering between Fermi arc states and track states. To disentangle these features in Fig.~\ref{fig:QPI_track_H3_McCormick}b and Fig.~\ref{fig:QPI_track_H3_McCormick}c we plot ``partial scattering integrals'', i.e., performing the integral in Eq.~(\ref{eq:QPIintegral}) only over some parts of the Brillouin zone. Thus, in Fig.~\ref{fig:QPI_track_H3_McCormick}b we  integrate over $\bs{q}$ only over the region $-\pi/2 < q_x < -\pi/4$, $-\pi/2 < q_z < \pi/2$, such that we only take into account the scattering of the states belonging to one Fermi arc with the rest of the BZ. We therefore eliminate the contributions to the total weight coming from scattering between the track states, and we are left with intra-Fermi-arc scattering, inter-Fermi-arc scattering, as well as scattering between Fermi-arc states and track states; these, however, can be easily identified based on their wavevectors. We mark each of these processes by black arrows in the figure. Similarly, in Fig.~\ref{fig:QPI_track_H3_McCormick}c we integrate over $\bs{q}$ only over the region $-\pi < q_x < -\pi/2$, $-\pi/2 < q_z < \pi/2$,  such that we only take into account the scattering of the states over one track state with the rest of the BZ. Each type of scattering processes is identified in the figure with the resulting features. 

Note that the intra-arc scattering is greatly suppressed  for this regime compared to the inter-arc scattering.
 
\section{Conclusion}
We have used a new analytical technique introduced in Ref.~[\onlinecite{Pinon2020}] and the T-matrix formalism to calculate the surface Green's functions as well as the QPI patterns for several different models describing Weyls semimetals. For the Kourtis and Lau models\cite{Kourtis2016,Lau2017} we found that no inter-arc interference patterns and correspondingly very weak interference patterns arise, in contrast to original predictions based on JDOS and SSP calculations. For the minimal McCormick model\cite{McCormick2017} both inter-arc and intra-arc scattering features are visible for type I models, as well as scattering between the track states and the Fermi-arc states and intra-track scattering for the type II models.

We stress once more that the technique proposed in Ref.~[\onlinecite{Pinon2020}] is very well suited to describe analytically the surface Green's functions for complex tight-binding bulk models, and it requires no iterative numerical calculations. As demonstrated here, the resulting Green's functions can very easily and straightforwardly be used to calculate the surface QPI features.

\begin{acknowledgments}
V.K. would like to acknowledge Jens Bardarson for stimulating discussions, the ERC Starting Grant No. 679722, the Roland Gustafsson foundation for theoretical physics, and the Karl Engvers foundation for financial support. 
\end{acknowledgments}

\bibliography{biblio_Tmatrix_Weyl_QPI}

\begin{thebibliography}{33}%
\makeatletter
\providecommand \@ifxundefined [1]{%
 \@ifx{#1\undefined}
}%
\providecommand \@ifnum [1]{%
 \ifnum #1\expandafter \@firstoftwo
 \else \expandafter \@secondoftwo
 \fi
}%
\providecommand \@ifx [1]{%
 \ifx #1\expandafter \@firstoftwo
 \else \expandafter \@secondoftwo
 \fi
}%
\providecommand \natexlab [1]{#1}%
\providecommand \enquote  [1]{``#1''}%
\providecommand \bibnamefont  [1]{#1}%
\providecommand \bibfnamefont [1]{#1}%
\providecommand \citenamefont [1]{#1}%
\providecommand \href@noop [0]{\@secondoftwo}%
\providecommand \href [0]{\begingroup \@sanitize@url \@href}%
\providecommand \@href[1]{\@@startlink{#1}\@@href}%
\providecommand \@@href[1]{\endgroup#1\@@endlink}%
\providecommand \@sanitize@url [0]{\catcode `\\12\catcode `\$12\catcode
  `\&12\catcode `\#12\catcode `\^12\catcode `\_12\catcode `\%12\relax}%
\providecommand \@@startlink[1]{}%
\providecommand \@@endlink[0]{}%
\providecommand \url  [0]{\begingroup\@sanitize@url \@url }%
\providecommand \@url [1]{\endgroup\@href {#1}{\urlprefix }}%
\providecommand \urlprefix  [0]{URL }%
\providecommand \Eprint [0]{\href }%
\providecommand \doibase [0]{http://dx.doi.org/}%
\providecommand \selectlanguage [0]{\@gobble}%
\providecommand \bibinfo  [0]{\@secondoftwo}%
\providecommand \bibfield  [0]{\@secondoftwo}%
\providecommand \translation [1]{[#1]}%
\providecommand \BibitemOpen [0]{}%
\providecommand \bibitemStop [0]{}%
\providecommand \bibitemNoStop [0]{.\EOS\space}%
\providecommand \EOS [0]{\spacefactor3000\relax}%
\providecommand \BibitemShut  [1]{\csname bibitem#1\endcsname}%
\let\auto@bib@innerbib\@empty
\bibitem [{\citenamefont {Turner}\ and\ \citenamefont
  {Vishwanath}(2013)}]{Turner2013}%
  \BibitemOpen
  \bibfield  {author} {\bibinfo {author} {\bibfnamefont {A.~M.}\ \bibnamefont
  {Turner}}\ and\ \bibinfo {author} {\bibfnamefont {A.}~\bibnamefont
  {Vishwanath}},\ }in\ \href {\doibase 10.1016/b978-0-444-63314-9.00011-1}
  {\emph {\bibinfo {booktitle} {Contemporary Concepts of Condensed Matter
  Science}}}\ (\bibinfo  {publisher} {Elsevier},\ \bibinfo {year} {2013})\ pp.\
  \bibinfo {pages} {293--324}\BibitemShut {NoStop}%
\bibitem [{\citenamefont {Mitchell}\ and\ \citenamefont
  {Fritz}(2016)}]{Mitchell2016}%
  \BibitemOpen
  \bibfield  {author} {\bibinfo {author} {\bibfnamefont {A.~K.}\ \bibnamefont
  {Mitchell}}\ and\ \bibinfo {author} {\bibfnamefont {L.}~\bibnamefont
  {Fritz}},\ }\href {\doibase 10.1103/PhysRevB.93.035137} {\bibfield  {journal}
  {\bibinfo  {journal} {Phys. Rev. B}\ }\textbf {\bibinfo {volume} {93}},\
  \bibinfo {pages} {035137} (\bibinfo {year} {2016})}\BibitemShut {NoStop}%
\bibitem [{\citenamefont {Gyenis}\ \emph {et~al.}(2016)\citenamefont {Gyenis},
  \citenamefont {Inoue}, \citenamefont {Jeon}, \citenamefont {Zhou},
  \citenamefont {Feldman}, \citenamefont {Wang}, \citenamefont {Li},
  \citenamefont {Jiang}, \citenamefont {Gibson}, \citenamefont {Kushwaha},
  \citenamefont {Krizan}, \citenamefont {Ni}, \citenamefont {Cava},
  \citenamefont {Bernevig},\ and\ \citenamefont {Yazdani}}]{Gyenis2016}%
  \BibitemOpen
  \bibfield  {author} {\bibinfo {author} {\bibfnamefont {A.}~\bibnamefont
  {Gyenis}}, \bibinfo {author} {\bibfnamefont {H.}~\bibnamefont {Inoue}},
  \bibinfo {author} {\bibfnamefont {S.}~\bibnamefont {Jeon}}, \bibinfo {author}
  {\bibfnamefont {B.~B.}\ \bibnamefont {Zhou}}, \bibinfo {author}
  {\bibfnamefont {B.~E.}\ \bibnamefont {Feldman}}, \bibinfo {author}
  {\bibfnamefont {Z.}~\bibnamefont {Wang}}, \bibinfo {author} {\bibfnamefont
  {J.}~\bibnamefont {Li}}, \bibinfo {author} {\bibfnamefont {S.}~\bibnamefont
  {Jiang}}, \bibinfo {author} {\bibfnamefont {Q.~D.}\ \bibnamefont {Gibson}},
  \bibinfo {author} {\bibfnamefont {S.~K.}\ \bibnamefont {Kushwaha}}, \bibinfo
  {author} {\bibfnamefont {J.~W.}\ \bibnamefont {Krizan}}, \bibinfo {author}
  {\bibfnamefont {N.}~\bibnamefont {Ni}}, \bibinfo {author} {\bibfnamefont
  {R.~J.}\ \bibnamefont {Cava}}, \bibinfo {author} {\bibfnamefont {B.~A.}\
  \bibnamefont {Bernevig}}, \ and\ \bibinfo {author} {\bibfnamefont
  {A.}~\bibnamefont {Yazdani}},\ }\href {\doibase
  10.1088/1367-2630/18/10/105003} {\bibfield  {journal} {\bibinfo  {journal}
  {New Journal of Physics}\ }\textbf {\bibinfo {volume} {18}},\ \bibinfo
  {pages} {105003} (\bibinfo {year} {2016})}\BibitemShut {NoStop}%
\bibitem [{\citenamefont {Lambert}\ \emph {et~al.}(2016)\citenamefont
  {Lambert}, \citenamefont {Schnyder}, \citenamefont {Moessner},\ and\
  \citenamefont {Eremin}}]{Lambert2016}%
  \BibitemOpen
  \bibfield  {author} {\bibinfo {author} {\bibfnamefont {F.}~\bibnamefont
  {Lambert}}, \bibinfo {author} {\bibfnamefont {A.~P.}\ \bibnamefont
  {Schnyder}}, \bibinfo {author} {\bibfnamefont {R.}~\bibnamefont {Moessner}},
  \ and\ \bibinfo {author} {\bibfnamefont {I.}~\bibnamefont {Eremin}},\ }\href
  {\doibase 10.1103/PhysRevB.94.165146} {\bibfield  {journal} {\bibinfo
  {journal} {Phys. Rev. B}\ }\textbf {\bibinfo {volume} {94}},\ \bibinfo
  {pages} {165146} (\bibinfo {year} {2016})}\BibitemShut {NoStop}%
\bibitem [{\citenamefont {Chang}\ \emph {et~al.}(2016)\citenamefont {Chang},
  \citenamefont {Xu}, \citenamefont {Zheng}, \citenamefont {Lee}, \citenamefont
  {Huang}, \citenamefont {Belopolski}, \citenamefont {Sanchez}, \citenamefont
  {Bian}, \citenamefont {Alidoust}, \citenamefont {Chang}, \citenamefont {Hsu},
  \citenamefont {Jeng}, \citenamefont {Bansil}, \citenamefont {Lin},\ and\
  \citenamefont {Hasan}}]{Chang2016}%
  \BibitemOpen
  \bibfield  {author} {\bibinfo {author} {\bibfnamefont {G.}~\bibnamefont
  {Chang}}, \bibinfo {author} {\bibfnamefont {S.-Y.}\ \bibnamefont {Xu}},
  \bibinfo {author} {\bibfnamefont {H.}~\bibnamefont {Zheng}}, \bibinfo
  {author} {\bibfnamefont {C.-C.}\ \bibnamefont {Lee}}, \bibinfo {author}
  {\bibfnamefont {S.-M.}\ \bibnamefont {Huang}}, \bibinfo {author}
  {\bibfnamefont {I.}~\bibnamefont {Belopolski}}, \bibinfo {author}
  {\bibfnamefont {D.~S.}\ \bibnamefont {Sanchez}}, \bibinfo {author}
  {\bibfnamefont {G.}~\bibnamefont {Bian}}, \bibinfo {author} {\bibfnamefont
  {N.}~\bibnamefont {Alidoust}}, \bibinfo {author} {\bibfnamefont {T.-R.}\
  \bibnamefont {Chang}}, \bibinfo {author} {\bibfnamefont {C.-H.}\ \bibnamefont
  {Hsu}}, \bibinfo {author} {\bibfnamefont {H.-T.}\ \bibnamefont {Jeng}},
  \bibinfo {author} {\bibfnamefont {A.}~\bibnamefont {Bansil}}, \bibinfo
  {author} {\bibfnamefont {H.}~\bibnamefont {Lin}}, \ and\ \bibinfo {author}
  {\bibfnamefont {M.~Z.}\ \bibnamefont {Hasan}},\ }\href {\doibase
  10.1103/PhysRevLett.116.066601} {\bibfield  {journal} {\bibinfo  {journal}
  {Phys. Rev. Lett.}\ }\textbf {\bibinfo {volume} {116}},\ \bibinfo {pages}
  {066601} (\bibinfo {year} {2016})}\BibitemShut {NoStop}%
\bibitem [{\citenamefont {Kourtis}\ \emph {et~al.}(2016)\citenamefont
  {Kourtis}, \citenamefont {Li}, \citenamefont {Wang}, \citenamefont
  {Yazdani},\ and\ \citenamefont {Bernevig}}]{Kourtis2016}%
  \BibitemOpen
  \bibfield  {author} {\bibinfo {author} {\bibfnamefont {S.}~\bibnamefont
  {Kourtis}}, \bibinfo {author} {\bibfnamefont {J.}~\bibnamefont {Li}},
  \bibinfo {author} {\bibfnamefont {Z.}~\bibnamefont {Wang}}, \bibinfo {author}
  {\bibfnamefont {A.}~\bibnamefont {Yazdani}}, \ and\ \bibinfo {author}
  {\bibfnamefont {B.~A.}\ \bibnamefont {Bernevig}},\ }\href {\doibase
  10.1103/PhysRevB.93.041109} {\bibfield  {journal} {\bibinfo  {journal} {Phys.
  Rev. B}\ }\textbf {\bibinfo {volume} {93}},\ \bibinfo {pages} {041109}
  (\bibinfo {year} {2016})}\BibitemShut {NoStop}%
\bibitem [{\citenamefont {McCormick}\ \emph {et~al.}(2017)\citenamefont
  {McCormick}, \citenamefont {Kimchi},\ and\ \citenamefont
  {Trivedi}}]{McCormick2017}%
  \BibitemOpen
  \bibfield  {author} {\bibinfo {author} {\bibfnamefont {T.~M.}\ \bibnamefont
  {McCormick}}, \bibinfo {author} {\bibfnamefont {I.}~\bibnamefont {Kimchi}}, \
  and\ \bibinfo {author} {\bibfnamefont {N.}~\bibnamefont {Trivedi}},\ }\href
  {\doibase 10.1103/PhysRevB.95.075133} {\bibfield  {journal} {\bibinfo
  {journal} {Phys. Rev. B}\ }\textbf {\bibinfo {volume} {95}},\ \bibinfo
  {pages} {075133} (\bibinfo {year} {2017})}\BibitemShut {NoStop}%
\bibitem [{\citenamefont {Lau}\ \emph {et~al.}(2017)\citenamefont {Lau},
  \citenamefont {Koepernik}, \citenamefont {van~den Brink},\ and\ \citenamefont
  {Ortix}}]{Lau2017}%
  \BibitemOpen
  \bibfield  {author} {\bibinfo {author} {\bibfnamefont {A.}~\bibnamefont
  {Lau}}, \bibinfo {author} {\bibfnamefont {K.}~\bibnamefont {Koepernik}},
  \bibinfo {author} {\bibfnamefont {J.}~\bibnamefont {van~den Brink}}, \ and\
  \bibinfo {author} {\bibfnamefont {C.}~\bibnamefont {Ortix}},\ }\href
  {\doibase 10.1103/PhysRevLett.119.076801} {\bibfield  {journal} {\bibinfo
  {journal} {Phys. Rev. Lett.}\ }\textbf {\bibinfo {volume} {119}},\ \bibinfo
  {pages} {076801} (\bibinfo {year} {2017})}\BibitemShut {NoStop}%
\bibitem [{\citenamefont {Xu}\ \emph {et~al.}(2018)\citenamefont {Xu},
  \citenamefont {Liu}, \citenamefont {Shi}, \citenamefont {Muechler},
  \citenamefont {Gayles}, \citenamefont {Felser},\ and\ \citenamefont
  {Sun}}]{Xu2018}%
  \BibitemOpen
  \bibfield  {author} {\bibinfo {author} {\bibfnamefont {Q.}~\bibnamefont
  {Xu}}, \bibinfo {author} {\bibfnamefont {E.}~\bibnamefont {Liu}}, \bibinfo
  {author} {\bibfnamefont {W.}~\bibnamefont {Shi}}, \bibinfo {author}
  {\bibfnamefont {L.}~\bibnamefont {Muechler}}, \bibinfo {author}
  {\bibfnamefont {J.}~\bibnamefont {Gayles}}, \bibinfo {author} {\bibfnamefont
  {C.}~\bibnamefont {Felser}}, \ and\ \bibinfo {author} {\bibfnamefont
  {Y.}~\bibnamefont {Sun}},\ }\href {\doibase 10.1103/PhysRevB.97.235416}
  {\bibfield  {journal} {\bibinfo  {journal} {Phys. Rev. B}\ }\textbf {\bibinfo
  {volume} {97}},\ \bibinfo {pages} {235416} (\bibinfo {year}
  {2018})}\BibitemShut {NoStop}%
\bibitem [{\citenamefont {Zheng}\ and\ \citenamefont
  {Hasan}(2018)}]{Zheng2018}%
  \BibitemOpen
  \bibfield  {author} {\bibinfo {author} {\bibfnamefont {H.}~\bibnamefont
  {Zheng}}\ and\ \bibinfo {author} {\bibfnamefont {M.~Z.}\ \bibnamefont
  {Hasan}},\ }\href {\doibase 10.1080/23746149.2018.1466661} {\bibfield
  {journal} {\bibinfo  {journal} {Advances in Physics: X}\ }\textbf {\bibinfo
  {volume} {3}},\ \bibinfo {pages} {1466661} (\bibinfo {year}
  {2018})}\BibitemShut {NoStop}%
\bibitem [{\citenamefont {Jeon}\ \emph {et~al.}(2014)\citenamefont {Jeon},
  \citenamefont {Zhou}, \citenamefont {Gyenis}, \citenamefont {Feldman},
  \citenamefont {Kimchi}, \citenamefont {Potter}, \citenamefont {Gibson},
  \citenamefont {Cava}, \citenamefont {Vishwanath},\ and\ \citenamefont
  {Yazdani}}]{Jeon2014}%
  \BibitemOpen
  \bibfield  {author} {\bibinfo {author} {\bibfnamefont {S.}~\bibnamefont
  {Jeon}}, \bibinfo {author} {\bibfnamefont {B.~B.}\ \bibnamefont {Zhou}},
  \bibinfo {author} {\bibfnamefont {A.}~\bibnamefont {Gyenis}}, \bibinfo
  {author} {\bibfnamefont {B.~E.}\ \bibnamefont {Feldman}}, \bibinfo {author}
  {\bibfnamefont {I.}~\bibnamefont {Kimchi}}, \bibinfo {author} {\bibfnamefont
  {A.~C.}\ \bibnamefont {Potter}}, \bibinfo {author} {\bibfnamefont {Q.~D.}\
  \bibnamefont {Gibson}}, \bibinfo {author} {\bibfnamefont {R.~J.}\
  \bibnamefont {Cava}}, \bibinfo {author} {\bibfnamefont {A.}~\bibnamefont
  {Vishwanath}}, \ and\ \bibinfo {author} {\bibfnamefont {A.}~\bibnamefont
  {Yazdani}},\ }\href {\doibase 10.1038/nmat4023} {\bibfield  {journal}
  {\bibinfo  {journal} {Nature Materials}\ }\textbf {\bibinfo {volume} {13}},\
  \bibinfo {pages} {851} (\bibinfo {year} {2014})}\BibitemShut {NoStop}%
\bibitem [{\citenamefont {Deng}\ \emph {et~al.}(2016)\citenamefont {Deng},
  \citenamefont {Wan}, \citenamefont {Deng}, \citenamefont {Zhang},
  \citenamefont {Ding}, \citenamefont {Wang}, \citenamefont {Yan},
  \citenamefont {Huang}, \citenamefont {Zhang}, \citenamefont {Xu},
  \citenamefont {Denlinger}, \citenamefont {Fedorov}, \citenamefont {Yang},
  \citenamefont {Duan}, \citenamefont {Yao}, \citenamefont {Wu}, \citenamefont
  {Fan}, \citenamefont {Zhang}, \citenamefont {Chen},\ and\ \citenamefont
  {Zhou}}]{Deng2016}%
  \BibitemOpen
  \bibfield  {author} {\bibinfo {author} {\bibfnamefont {K.}~\bibnamefont
  {Deng}}, \bibinfo {author} {\bibfnamefont {G.}~\bibnamefont {Wan}}, \bibinfo
  {author} {\bibfnamefont {P.}~\bibnamefont {Deng}}, \bibinfo {author}
  {\bibfnamefont {K.}~\bibnamefont {Zhang}}, \bibinfo {author} {\bibfnamefont
  {S.}~\bibnamefont {Ding}}, \bibinfo {author} {\bibfnamefont {E.}~\bibnamefont
  {Wang}}, \bibinfo {author} {\bibfnamefont {M.}~\bibnamefont {Yan}}, \bibinfo
  {author} {\bibfnamefont {H.}~\bibnamefont {Huang}}, \bibinfo {author}
  {\bibfnamefont {H.}~\bibnamefont {Zhang}}, \bibinfo {author} {\bibfnamefont
  {Z.}~\bibnamefont {Xu}}, \bibinfo {author} {\bibfnamefont {J.}~\bibnamefont
  {Denlinger}}, \bibinfo {author} {\bibfnamefont {A.}~\bibnamefont {Fedorov}},
  \bibinfo {author} {\bibfnamefont {H.}~\bibnamefont {Yang}}, \bibinfo {author}
  {\bibfnamefont {W.}~\bibnamefont {Duan}}, \bibinfo {author} {\bibfnamefont
  {H.}~\bibnamefont {Yao}}, \bibinfo {author} {\bibfnamefont {Y.}~\bibnamefont
  {Wu}}, \bibinfo {author} {\bibfnamefont {S.}~\bibnamefont {Fan}}, \bibinfo
  {author} {\bibfnamefont {H.}~\bibnamefont {Zhang}}, \bibinfo {author}
  {\bibfnamefont {X.}~\bibnamefont {Chen}}, \ and\ \bibinfo {author}
  {\bibfnamefont {S.}~\bibnamefont {Zhou}},\ }\href {\doibase
  10.1038/nphys3871} {\bibfield  {journal} {\bibinfo  {journal} {Nature
  Physics}\ }\textbf {\bibinfo {volume} {12}},\ \bibinfo {pages} {1105}
  (\bibinfo {year} {2016})}\BibitemShut {NoStop}%
\bibitem [{\citenamefont {Zheng}\ \emph
  {et~al.}(2016{\natexlab{a}})\citenamefont {Zheng}, \citenamefont {Xu},
  \citenamefont {Bian}, \citenamefont {Guo}, \citenamefont {Chang},
  \citenamefont {Sanchez}, \citenamefont {Belopolski}, \citenamefont {Lee},
  \citenamefont {Huang}, \citenamefont {Zhang}, \citenamefont {Sankar},
  \citenamefont {Alidoust}, \citenamefont {Chang}, \citenamefont {Wu},
  \citenamefont {Neupert}, \citenamefont {Chou}, \citenamefont {Jeng},
  \citenamefont {Yao}, \citenamefont {Bansil}, \citenamefont {Jia},
  \citenamefont {Lin},\ and\ \citenamefont {Hasan}}]{Zheng2016a}%
  \BibitemOpen
  \bibfield  {author} {\bibinfo {author} {\bibfnamefont {H.}~\bibnamefont
  {Zheng}}, \bibinfo {author} {\bibfnamefont {S.-Y.}\ \bibnamefont {Xu}},
  \bibinfo {author} {\bibfnamefont {G.}~\bibnamefont {Bian}}, \bibinfo {author}
  {\bibfnamefont {C.}~\bibnamefont {Guo}}, \bibinfo {author} {\bibfnamefont
  {G.}~\bibnamefont {Chang}}, \bibinfo {author} {\bibfnamefont {D.~S.}\
  \bibnamefont {Sanchez}}, \bibinfo {author} {\bibfnamefont {I.}~\bibnamefont
  {Belopolski}}, \bibinfo {author} {\bibfnamefont {C.-C.}\ \bibnamefont {Lee}},
  \bibinfo {author} {\bibfnamefont {S.-M.}\ \bibnamefont {Huang}}, \bibinfo
  {author} {\bibfnamefont {X.}~\bibnamefont {Zhang}}, \bibinfo {author}
  {\bibfnamefont {R.}~\bibnamefont {Sankar}}, \bibinfo {author} {\bibfnamefont
  {N.}~\bibnamefont {Alidoust}}, \bibinfo {author} {\bibfnamefont {T.-R.}\
  \bibnamefont {Chang}}, \bibinfo {author} {\bibfnamefont {F.}~\bibnamefont
  {Wu}}, \bibinfo {author} {\bibfnamefont {T.}~\bibnamefont {Neupert}},
  \bibinfo {author} {\bibfnamefont {F.}~\bibnamefont {Chou}}, \bibinfo {author}
  {\bibfnamefont {H.-T.}\ \bibnamefont {Jeng}}, \bibinfo {author}
  {\bibfnamefont {N.}~\bibnamefont {Yao}}, \bibinfo {author} {\bibfnamefont
  {A.}~\bibnamefont {Bansil}}, \bibinfo {author} {\bibfnamefont
  {S.}~\bibnamefont {Jia}}, \bibinfo {author} {\bibfnamefont {H.}~\bibnamefont
  {Lin}}, \ and\ \bibinfo {author} {\bibfnamefont {M.~Z.}\ \bibnamefont
  {Hasan}},\ }\href {\doibase 10.1021/acsnano.5b06807} {\bibfield  {journal}
  {\bibinfo  {journal} {{ACS} Nano}\ }\textbf {\bibinfo {volume} {10}},\
  \bibinfo {pages} {1378} (\bibinfo {year} {2016}{\natexlab{a}})}\BibitemShut
  {NoStop}%
\bibitem [{\citenamefont {Zheng}\ \emph
  {et~al.}(2016{\natexlab{b}})\citenamefont {Zheng}, \citenamefont {Bian},
  \citenamefont {Chang}, \citenamefont {Lu}, \citenamefont {Xu}, \citenamefont
  {Wang}, \citenamefont {Chang}, \citenamefont {Zhang}, \citenamefont
  {Belopolski}, \citenamefont {Alidoust}, \citenamefont {Sanchez},
  \citenamefont {Song}, \citenamefont {Jeng}, \citenamefont {Yao},
  \citenamefont {Bansil}, \citenamefont {Jia}, \citenamefont {Lin},\ and\
  \citenamefont {Hasan}}]{Zheng2016b}%
  \BibitemOpen
  \bibfield  {author} {\bibinfo {author} {\bibfnamefont {H.}~\bibnamefont
  {Zheng}}, \bibinfo {author} {\bibfnamefont {G.}~\bibnamefont {Bian}},
  \bibinfo {author} {\bibfnamefont {G.}~\bibnamefont {Chang}}, \bibinfo
  {author} {\bibfnamefont {H.}~\bibnamefont {Lu}}, \bibinfo {author}
  {\bibfnamefont {S.-Y.}\ \bibnamefont {Xu}}, \bibinfo {author} {\bibfnamefont
  {G.}~\bibnamefont {Wang}}, \bibinfo {author} {\bibfnamefont {T.-R.}\
  \bibnamefont {Chang}}, \bibinfo {author} {\bibfnamefont {S.}~\bibnamefont
  {Zhang}}, \bibinfo {author} {\bibfnamefont {I.}~\bibnamefont {Belopolski}},
  \bibinfo {author} {\bibfnamefont {N.}~\bibnamefont {Alidoust}}, \bibinfo
  {author} {\bibfnamefont {D.~S.}\ \bibnamefont {Sanchez}}, \bibinfo {author}
  {\bibfnamefont {F.}~\bibnamefont {Song}}, \bibinfo {author} {\bibfnamefont
  {H.-T.}\ \bibnamefont {Jeng}}, \bibinfo {author} {\bibfnamefont
  {N.}~\bibnamefont {Yao}}, \bibinfo {author} {\bibfnamefont {A.}~\bibnamefont
  {Bansil}}, \bibinfo {author} {\bibfnamefont {S.}~\bibnamefont {Jia}},
  \bibinfo {author} {\bibfnamefont {H.}~\bibnamefont {Lin}}, \ and\ \bibinfo
  {author} {\bibfnamefont {M.~Z.}\ \bibnamefont {Hasan}},\ }\href {\doibase
  10.1103/PhysRevLett.117.266804} {\bibfield  {journal} {\bibinfo  {journal}
  {Phys. Rev. Lett.}\ }\textbf {\bibinfo {volume} {117}},\ \bibinfo {pages}
  {266804} (\bibinfo {year} {2016}{\natexlab{b}})}\BibitemShut {NoStop}%
\bibitem [{\citenamefont {Inoue}\ \emph {et~al.}(2016)\citenamefont {Inoue},
  \citenamefont {Gyenis}, \citenamefont {Wang}, \citenamefont {Li},
  \citenamefont {Oh}, \citenamefont {Jiang}, \citenamefont {Ni}, \citenamefont
  {Bernevig},\ and\ \citenamefont {Yazdani}}]{Inoue2016}%
  \BibitemOpen
  \bibfield  {author} {\bibinfo {author} {\bibfnamefont {H.}~\bibnamefont
  {Inoue}}, \bibinfo {author} {\bibfnamefont {A.}~\bibnamefont {Gyenis}},
  \bibinfo {author} {\bibfnamefont {Z.}~\bibnamefont {Wang}}, \bibinfo {author}
  {\bibfnamefont {J.}~\bibnamefont {Li}}, \bibinfo {author} {\bibfnamefont
  {S.~W.}\ \bibnamefont {Oh}}, \bibinfo {author} {\bibfnamefont
  {S.}~\bibnamefont {Jiang}}, \bibinfo {author} {\bibfnamefont
  {N.}~\bibnamefont {Ni}}, \bibinfo {author} {\bibfnamefont {B.~A.}\
  \bibnamefont {Bernevig}}, \ and\ \bibinfo {author} {\bibfnamefont
  {A.}~\bibnamefont {Yazdani}},\ }\href {\doibase 10.1126/science.aad8766}
  {\bibfield  {journal} {\bibinfo  {journal} {Science}\ }\textbf {\bibinfo
  {volume} {351}},\ \bibinfo {pages} {1184} (\bibinfo {year}
  {2016})}\BibitemShut {NoStop}%
\bibitem [{\citenamefont {Batabyal}\ \emph {et~al.}(2016)\citenamefont
  {Batabyal}, \citenamefont {Morali}, \citenamefont {Avraham}, \citenamefont
  {Sun}, \citenamefont {Schmidt}, \citenamefont {Felser}, \citenamefont
  {Stern}, \citenamefont {Yan},\ and\ \citenamefont
  {Beidenkopf}}]{Batabyal2016}%
  \BibitemOpen
  \bibfield  {author} {\bibinfo {author} {\bibfnamefont {R.}~\bibnamefont
  {Batabyal}}, \bibinfo {author} {\bibfnamefont {N.}~\bibnamefont {Morali}},
  \bibinfo {author} {\bibfnamefont {N.}~\bibnamefont {Avraham}}, \bibinfo
  {author} {\bibfnamefont {Y.}~\bibnamefont {Sun}}, \bibinfo {author}
  {\bibfnamefont {M.}~\bibnamefont {Schmidt}}, \bibinfo {author} {\bibfnamefont
  {C.}~\bibnamefont {Felser}}, \bibinfo {author} {\bibfnamefont
  {A.}~\bibnamefont {Stern}}, \bibinfo {author} {\bibfnamefont
  {B.}~\bibnamefont {Yan}}, \ and\ \bibinfo {author} {\bibfnamefont
  {H.}~\bibnamefont {Beidenkopf}},\ }\href {\doibase 10.1126/sciadv.1600709}
  {\bibfield  {journal} {\bibinfo  {journal} {Science Advances}\ }\textbf
  {\bibinfo {volume} {2}},\ \bibinfo {pages} {e1600709} (\bibinfo {year}
  {2016})}\BibitemShut {NoStop}%
\bibitem [{\citenamefont {Zhang}\ \emph {et~al.}(2017)\citenamefont {Zhang},
  \citenamefont {Wu}, \citenamefont {Zhang}, \citenamefont {Cheong},
  \citenamefont {Soluyanov},\ and\ \citenamefont {Wu}}]{Zhang2017}%
  \BibitemOpen
  \bibfield  {author} {\bibinfo {author} {\bibfnamefont {W.}~\bibnamefont
  {Zhang}}, \bibinfo {author} {\bibfnamefont {Q.}~\bibnamefont {Wu}}, \bibinfo
  {author} {\bibfnamefont {L.}~\bibnamefont {Zhang}}, \bibinfo {author}
  {\bibfnamefont {S.-W.}\ \bibnamefont {Cheong}}, \bibinfo {author}
  {\bibfnamefont {A.~A.}\ \bibnamefont {Soluyanov}}, \ and\ \bibinfo {author}
  {\bibfnamefont {W.}~\bibnamefont {Wu}},\ }\href {\doibase
  10.1103/PhysRevB.96.165125} {\bibfield  {journal} {\bibinfo  {journal} {Phys.
  Rev. B}\ }\textbf {\bibinfo {volume} {96}},\ \bibinfo {pages} {165125}
  (\bibinfo {year} {2017})}\BibitemShut {NoStop}%
\bibitem [{\citenamefont {Deng}\ \emph {et~al.}(2017)\citenamefont {Deng},
  \citenamefont {Xu}, \citenamefont {Deng}, \citenamefont {Zhang},
  \citenamefont {Wu}, \citenamefont {Zhang}, \citenamefont {Zhou},\ and\
  \citenamefont {Chen}}]{Deng2017}%
  \BibitemOpen
  \bibfield  {author} {\bibinfo {author} {\bibfnamefont {P.}~\bibnamefont
  {Deng}}, \bibinfo {author} {\bibfnamefont {Z.}~\bibnamefont {Xu}}, \bibinfo
  {author} {\bibfnamefont {K.}~\bibnamefont {Deng}}, \bibinfo {author}
  {\bibfnamefont {K.}~\bibnamefont {Zhang}}, \bibinfo {author} {\bibfnamefont
  {Y.}~\bibnamefont {Wu}}, \bibinfo {author} {\bibfnamefont {H.}~\bibnamefont
  {Zhang}}, \bibinfo {author} {\bibfnamefont {S.}~\bibnamefont {Zhou}}, \ and\
  \bibinfo {author} {\bibfnamefont {X.}~\bibnamefont {Chen}},\ }\href {\doibase
  10.1103/PhysRevB.95.245110} {\bibfield  {journal} {\bibinfo  {journal} {Phys.
  Rev. B}\ }\textbf {\bibinfo {volume} {95}},\ \bibinfo {pages} {245110}
  (\bibinfo {year} {2017})}\BibitemShut {NoStop}%
\bibitem [{\citenamefont {Lin}\ \emph {et~al.}(2017)\citenamefont {Lin},
  \citenamefont {Arafune}, \citenamefont {Liu}, \citenamefont {Yoshimura},
  \citenamefont {Feng}, \citenamefont {Kawahara}, \citenamefont {Ni},
  \citenamefont {Minamitani}, \citenamefont {Watanabe}, \citenamefont {Shi},
  \citenamefont {Kawai}, \citenamefont {Chiang}, \citenamefont {Matsuda},\ and\
  \citenamefont {Takagi}}]{Lin2017}%
  \BibitemOpen
  \bibfield  {author} {\bibinfo {author} {\bibfnamefont {C.-L.}\ \bibnamefont
  {Lin}}, \bibinfo {author} {\bibfnamefont {R.}~\bibnamefont {Arafune}},
  \bibinfo {author} {\bibfnamefont {R.-Y.}\ \bibnamefont {Liu}}, \bibinfo
  {author} {\bibfnamefont {M.}~\bibnamefont {Yoshimura}}, \bibinfo {author}
  {\bibfnamefont {B.}~\bibnamefont {Feng}}, \bibinfo {author} {\bibfnamefont
  {K.}~\bibnamefont {Kawahara}}, \bibinfo {author} {\bibfnamefont
  {Z.}~\bibnamefont {Ni}}, \bibinfo {author} {\bibfnamefont {E.}~\bibnamefont
  {Minamitani}}, \bibinfo {author} {\bibfnamefont {S.}~\bibnamefont
  {Watanabe}}, \bibinfo {author} {\bibfnamefont {Y.}~\bibnamefont {Shi}},
  \bibinfo {author} {\bibfnamefont {M.}~\bibnamefont {Kawai}}, \bibinfo
  {author} {\bibfnamefont {T.-C.}\ \bibnamefont {Chiang}}, \bibinfo {author}
  {\bibfnamefont {I.}~\bibnamefont {Matsuda}}, \ and\ \bibinfo {author}
  {\bibfnamefont {N.}~\bibnamefont {Takagi}},\ }\href {\doibase
  10.1021/acsnano.7b06179} {\bibfield  {journal} {\bibinfo  {journal} {{ACS}
  Nano}\ }\textbf {\bibinfo {volume} {11}},\ \bibinfo {pages} {11459} (\bibinfo
  {year} {2017})}\BibitemShut {NoStop}%
\bibitem [{\citenamefont {Yuan}\ \emph {et~al.}(2018)\citenamefont {Yuan},
  \citenamefont {Yang}, \citenamefont {Peng}, \citenamefont {Wang},
  \citenamefont {Li}, \citenamefont {Yi}, \citenamefont {Xian}, \citenamefont
  {Shi},\ and\ \citenamefont {Fu}}]{Yuan2018}%
  \BibitemOpen
  \bibfield  {author} {\bibinfo {author} {\bibfnamefont {Y.}~\bibnamefont
  {Yuan}}, \bibinfo {author} {\bibfnamefont {X.}~\bibnamefont {Yang}}, \bibinfo
  {author} {\bibfnamefont {L.}~\bibnamefont {Peng}}, \bibinfo {author}
  {\bibfnamefont {Z.-J.}\ \bibnamefont {Wang}}, \bibinfo {author}
  {\bibfnamefont {J.}~\bibnamefont {Li}}, \bibinfo {author} {\bibfnamefont
  {C.-J.}\ \bibnamefont {Yi}}, \bibinfo {author} {\bibfnamefont {J.-J.}\
  \bibnamefont {Xian}}, \bibinfo {author} {\bibfnamefont {Y.-G.}\ \bibnamefont
  {Shi}}, \ and\ \bibinfo {author} {\bibfnamefont {Y.-S.}\ \bibnamefont {Fu}},\
  }\href {\doibase 10.1103/PhysRevB.97.165435} {\bibfield  {journal} {\bibinfo
  {journal} {Phys. Rev. B}\ }\textbf {\bibinfo {volume} {97}},\ \bibinfo
  {pages} {165435} (\bibinfo {year} {2018})}\BibitemShut {NoStop}%
\bibitem [{\citenamefont {Lin}\ \emph {et~al.}(2018)\citenamefont {Lin},
  \citenamefont {Arafune}, \citenamefont {Minamitani}, \citenamefont {Kawai},\
  and\ \citenamefont {Takagi}}]{Lin2018}%
  \BibitemOpen
  \bibfield  {author} {\bibinfo {author} {\bibfnamefont {C.-L.}\ \bibnamefont
  {Lin}}, \bibinfo {author} {\bibfnamefont {R.}~\bibnamefont {Arafune}},
  \bibinfo {author} {\bibfnamefont {E.}~\bibnamefont {Minamitani}}, \bibinfo
  {author} {\bibfnamefont {M.}~\bibnamefont {Kawai}}, \ and\ \bibinfo {author}
  {\bibfnamefont {N.}~\bibnamefont {Takagi}},\ }\href {\doibase
  10.1088/1361-648x/aaab95} {\bibfield  {journal} {\bibinfo  {journal} {Journal
  of Physics: Condensed Matter}\ }\textbf {\bibinfo {volume} {30}},\ \bibinfo
  {pages} {105703} (\bibinfo {year} {2018})}\BibitemShut {NoStop}%
\bibitem [{\citenamefont {Morali}\ \emph {et~al.}(2019)\citenamefont {Morali},
  \citenamefont {Batabyal}, \citenamefont {Nag}, \citenamefont {Liu},
  \citenamefont {Xu}, \citenamefont {Sun}, \citenamefont {Yan}, \citenamefont
  {Felser}, \citenamefont {Avraham},\ and\ \citenamefont
  {Beidenkopf}}]{Morali2019}%
  \BibitemOpen
  \bibfield  {author} {\bibinfo {author} {\bibfnamefont {N.}~\bibnamefont
  {Morali}}, \bibinfo {author} {\bibfnamefont {R.}~\bibnamefont {Batabyal}},
  \bibinfo {author} {\bibfnamefont {P.~K.}\ \bibnamefont {Nag}}, \bibinfo
  {author} {\bibfnamefont {E.}~\bibnamefont {Liu}}, \bibinfo {author}
  {\bibfnamefont {Q.}~\bibnamefont {Xu}}, \bibinfo {author} {\bibfnamefont
  {Y.}~\bibnamefont {Sun}}, \bibinfo {author} {\bibfnamefont {B.}~\bibnamefont
  {Yan}}, \bibinfo {author} {\bibfnamefont {C.}~\bibnamefont {Felser}},
  \bibinfo {author} {\bibfnamefont {N.}~\bibnamefont {Avraham}}, \ and\
  \bibinfo {author} {\bibfnamefont {H.}~\bibnamefont {Beidenkopf}},\ }\href
  {\doibase 10.1126/science.aav2334} {\bibfield  {journal} {\bibinfo  {journal}
  {Science}\ }\textbf {\bibinfo {volume} {365}},\ \bibinfo {pages} {1286}
  (\bibinfo {year} {2019})}\BibitemShut {NoStop}%
\bibitem [{\citenamefont {Sessi}\ \emph {et~al.}(2020)\citenamefont {Sessi},
  \citenamefont {Fan}, \citenamefont {Küster}, \citenamefont {Manna},
  \citenamefont {Schröter}, \citenamefont {Ji}, \citenamefont {Stolz},
  \citenamefont {Krieger}, \citenamefont {Pei}, \citenamefont {Kim},
  \citenamefont {Dudin}, \citenamefont {Cacho}, \citenamefont {Widmer},
  \citenamefont {Borrmann}, \citenamefont {Shi}, \citenamefont {Chang},
  \citenamefont {Sun}, \citenamefont {Felser},\ and\ \citenamefont
  {Parkin}}]{Sessi2020}%
  \BibitemOpen
  \bibfield  {author} {\bibinfo {author} {\bibfnamefont {P.}~\bibnamefont
  {Sessi}}, \bibinfo {author} {\bibfnamefont {F.-R.}\ \bibnamefont {Fan}},
  \bibinfo {author} {\bibfnamefont {F.}~\bibnamefont {Küster}}, \bibinfo
  {author} {\bibfnamefont {K.}~\bibnamefont {Manna}}, \bibinfo {author}
  {\bibfnamefont {N.~B.}\ \bibnamefont {Schröter}}, \bibinfo {author}
  {\bibfnamefont {J.-R.}\ \bibnamefont {Ji}}, \bibinfo {author} {\bibfnamefont
  {S.}~\bibnamefont {Stolz}}, \bibinfo {author} {\bibfnamefont {J.~A.}\
  \bibnamefont {Krieger}}, \bibinfo {author} {\bibfnamefont {D.}~\bibnamefont
  {Pei}}, \bibinfo {author} {\bibfnamefont {T.~K.}\ \bibnamefont {Kim}},
  \bibinfo {author} {\bibfnamefont {P.}~\bibnamefont {Dudin}}, \bibinfo
  {author} {\bibfnamefont {C.}~\bibnamefont {Cacho}}, \bibinfo {author}
  {\bibfnamefont {R.}~\bibnamefont {Widmer}}, \bibinfo {author} {\bibfnamefont
  {H.}~\bibnamefont {Borrmann}}, \bibinfo {author} {\bibfnamefont
  {W.}~\bibnamefont {Shi}}, \bibinfo {author} {\bibfnamefont {K.}~\bibnamefont
  {Chang}}, \bibinfo {author} {\bibfnamefont {Y.}~\bibnamefont {Sun}}, \bibinfo
  {author} {\bibfnamefont {C.}~\bibnamefont {Felser}}, \ and\ \bibinfo {author}
  {\bibfnamefont {S.~S.}\ \bibnamefont {Parkin}},\ }\href {\doibase
  arXiv:2005.03116} {\bibfield  {journal} {\bibinfo  {journal} {arXiv}\ }
  (\bibinfo {year} {2020}),\ arXiv:2005.03116}\BibitemShut {NoStop}%
\bibitem [{\citenamefont {Ziegler}\ \emph {et~al.}(1996)\citenamefont
  {Ziegler}, \citenamefont {Poilblanc}, \citenamefont {Preuss}, \citenamefont
  {Hanke},\ and\ \citenamefont {Scalapino}}]{Ziegler1996}%
  \BibitemOpen
  \bibfield  {author} {\bibinfo {author} {\bibfnamefont {W.}~\bibnamefont
  {Ziegler}}, \bibinfo {author} {\bibfnamefont {D.}~\bibnamefont {Poilblanc}},
  \bibinfo {author} {\bibfnamefont {R.}~\bibnamefont {Preuss}}, \bibinfo
  {author} {\bibfnamefont {W.}~\bibnamefont {Hanke}}, \ and\ \bibinfo {author}
  {\bibfnamefont {D.~J.}\ \bibnamefont {Scalapino}},\ }\href {\doibase
  10.1103/PhysRevB.53.8704} {\bibfield  {journal} {\bibinfo  {journal} {Phys.
  Rev. B}\ }\textbf {\bibinfo {volume} {53}},\ \bibinfo {pages} {8704}
  (\bibinfo {year} {1996})}\BibitemShut {NoStop}%
\bibitem [{\citenamefont {Salkola}\ \emph {et~al.}(1996)\citenamefont
  {Salkola}, \citenamefont {Balatsky},\ and\ \citenamefont
  {Scalapino}}]{Salkola1996}%
  \BibitemOpen
  \bibfield  {author} {\bibinfo {author} {\bibfnamefont {M.~I.}\ \bibnamefont
  {Salkola}}, \bibinfo {author} {\bibfnamefont {A.~V.}\ \bibnamefont
  {Balatsky}}, \ and\ \bibinfo {author} {\bibfnamefont {D.~J.}\ \bibnamefont
  {Scalapino}},\ }\href {\doibase 10.1103/PhysRevLett.77.1841} {\bibfield
  {journal} {\bibinfo  {journal} {Phys. Rev. Lett.}\ }\textbf {\bibinfo
  {volume} {77}},\ \bibinfo {pages} {1841} (\bibinfo {year}
  {1996})}\BibitemShut {NoStop}%
\bibitem [{\citenamefont {Mahan}(2000)}]{Mahan2000}%
  \BibitemOpen
  \bibfield  {author} {\bibinfo {author} {\bibfnamefont {G.~D.}\ \bibnamefont
  {Mahan}},\ }\href {\doibase 10.1007/978-1-4757-5714-9} {\emph {\bibinfo
  {title} {Many-Particle Physics}}}\ (\bibinfo  {publisher} {Springer {US}},\
  \bibinfo {year} {2000})\BibitemShut {NoStop}%
\bibitem [{\citenamefont {Bena}(2016)}]{Bena2016}%
  \BibitemOpen
  \bibfield  {author} {\bibinfo {author} {\bibfnamefont {C.}~\bibnamefont
  {Bena}},\ }\href {\doibase 10.1016/j.crhy.2015.11.006} {\bibfield  {journal}
  {\bibinfo  {journal} {Comptes Rendus Physique}\ }\textbf {\bibinfo {volume}
  {17}},\ \bibinfo {pages} {302} (\bibinfo {year} {2016})}\BibitemShut
  {NoStop}%
\bibitem [{\citenamefont {Pinon}\ \emph {et~al.}(2020)\citenamefont {Pinon},
  \citenamefont {Kaladzhyan},\ and\ \citenamefont {Bena}}]{Pinon2020}%
  \BibitemOpen
  \bibfield  {author} {\bibinfo {author} {\bibfnamefont {S.}~\bibnamefont
  {Pinon}}, \bibinfo {author} {\bibfnamefont {V.}~\bibnamefont {Kaladzhyan}}, \
  and\ \bibinfo {author} {\bibfnamefont {C.}~\bibnamefont {Bena}},\ }\href
  {\doibase 10.1103/PhysRevB.101.115405} {\bibfield  {journal} {\bibinfo
  {journal} {Phys. Rev. B}\ }\textbf {\bibinfo {volume} {101}},\ \bibinfo
  {pages} {115405} (\bibinfo {year} {2020})}\BibitemShut {NoStop}%
\bibitem [{Note1()}]{Note1}%
  \BibitemOpen
  \bibinfo {note} {Note that the impurity needs to contain a number of planes
  equal to the unit cell number of planes in the model, or more in case of
  higher-order hopping between planes}\BibitemShut {NoStop}%
\bibitem [{\citenamefont {Balatsky}\ \emph {et~al.}(2006)\citenamefont
  {Balatsky}, \citenamefont {Vekhter},\ and\ \citenamefont
  {Zhu}}]{Balatsky2006}%
  \BibitemOpen
  \bibfield  {author} {\bibinfo {author} {\bibfnamefont {A.~V.}\ \bibnamefont
  {Balatsky}}, \bibinfo {author} {\bibfnamefont {I.}~\bibnamefont {Vekhter}}, \
  and\ \bibinfo {author} {\bibfnamefont {J.-X.}\ \bibnamefont {Zhu}},\ }\href
  {\doibase 10.1103/RevModPhys.78.373} {\bibfield  {journal} {\bibinfo
  {journal} {Rev. Mod. Phys.}\ }\textbf {\bibinfo {volume} {78}},\ \bibinfo
  {pages} {373} (\bibinfo {year} {2006})}\BibitemShut {NoStop}%
\bibitem [{\citenamefont {Bena}(2008)}]{Bena2008}%
  \BibitemOpen
  \bibfield  {author} {\bibinfo {author} {\bibfnamefont {C.}~\bibnamefont
  {Bena}},\ }\href {\doibase 10.1103/PhysRevLett.100.076601} {\bibfield
  {journal} {\bibinfo  {journal} {Phys. Rev. Lett.}\ }\textbf {\bibinfo
  {volume} {100}},\ \bibinfo {pages} {076601} (\bibinfo {year}
  {2008})}\BibitemShut {NoStop}%
\bibitem [{\citenamefont {Kaladzhyan}\ \emph {et~al.}(2016)\citenamefont
  {Kaladzhyan}, \citenamefont {Simon},\ and\ \citenamefont
  {Bena}}]{Kaladzhyan2016}%
  \BibitemOpen
  \bibfield  {author} {\bibinfo {author} {\bibfnamefont {V.}~\bibnamefont
  {Kaladzhyan}}, \bibinfo {author} {\bibfnamefont {P.}~\bibnamefont {Simon}}, \
  and\ \bibinfo {author} {\bibfnamefont {C.}~\bibnamefont {Bena}},\ }\href
  {\doibase 10.1103/PhysRevB.94.134511} {\bibfield  {journal} {\bibinfo
  {journal} {Phys. Rev. B}\ }\textbf {\bibinfo {volume} {94}},\ \bibinfo
  {pages} {134511} (\bibinfo {year} {2016})}\BibitemShut {NoStop}%
\bibitem [{\citenamefont {Derry}\ \emph {et~al.}(2015)\citenamefont {Derry},
  \citenamefont {Mitchell},\ and\ \citenamefont {Logan}}]{Derry2015}%
  \BibitemOpen
  \bibfield  {author} {\bibinfo {author} {\bibfnamefont {P.~G.}\ \bibnamefont
  {Derry}}, \bibinfo {author} {\bibfnamefont {A.~K.}\ \bibnamefont {Mitchell}},
  \ and\ \bibinfo {author} {\bibfnamefont {D.~E.}\ \bibnamefont {Logan}},\
  }\href {\doibase 10.1103/PhysRevB.92.035126} {\bibfield  {journal} {\bibinfo
  {journal} {Phys. Rev. B}\ }\textbf {\bibinfo {volume} {92}},\ \bibinfo
  {pages} {035126} (\bibinfo {year} {2015})}\BibitemShut {NoStop}%
\end{thebibliography}%

\end{document}